\documentclass[letterpaper]{article} % DO NOT CHANGE THIS
\usepackage{aaai2026}  % DO NOT CHANGE THIS
\usepackage{times}  % DO NOT CHANGE THIS
\usepackage{helvet}  % DO NOT CHANGE THIS
\usepackage{courier}  % DO NOT CHANGE THIS
\usepackage[hyphens]{url}  % DO NOT CHANGE THIS
\usepackage{graphicx} % DO NOT CHANGE THIS
\usepackage{natbib}  % DO NOT CHANGE THIS AND DO NOT ADD ANY OPTIONS TO IT
\usepackage{caption} % DO NOT CHANGE THIS AND DO NOT ADD ANY OPTIONS TO IT
\usepackage{algorithm}
\usepackage{algorithmic}
\usepackage{booktabs}
\usepackage{makecell}
\usepackage{subcaption}

\usepackage{newfloat}
\usepackage{listings}
\DeclareCaptionStyle{ruled}{labelfont=normalfont,labelsep=colon,strut=off} % DO NOT CHANGE THIS
\floatstyle{ruled}
\newfloat{listing}{tb}{lst}{}
\floatname{listing}{Listing}
\title{Playing Games with My Heart: An Evaluation of AI Companion Apps}
\author {
    Maribeth Rauh\textsuperscript{\rm 1,},
    Dick A. H. Blankvoort\textsuperscript{\rm 1},
    Matias Duran\textsuperscript{\rm 2},
    Caoilfhionn Ní Dheoráin\textsuperscript{\rm 2},
    Harshvardhan J. Pandit\textsuperscript{\rm 1},
    Syrine Enneifer\textsuperscript{\rm 1, \rm 3},
    Siddharth Jaiswal\textsuperscript{\rm 4},
    Anthony Ventresque\textsuperscript{\rm 2},
    Abeba Birhane\textsuperscript{\rm 1}
}
\affiliations {
    \textsuperscript{\rm 1}AI Accountability Lab, School of Computer Science and Statistics (SCSS), Trinity College Dublin\\
    \textsuperscript{\rm 2}Complex Software Lab, SCSS, Trinity College Dublin\\
    \textsuperscript{\rm 3}Sapienza Università di Roma, DIAG\\
    \textsuperscript{\rm 4}Indian Institute of Technology Kharagpur\\
}
\begin{document}

\maketitle

\begin{abstract}
  The use of chatbots for various forms of companionship is growing rapidly, raising a myriad of questions about simulated relationships, emotional dependence, and psychological harm.
  While major platforms such as ChatGPT, Grok, and Character.AI are the subject of a growing body of research and legal inquiries, apps explicitly built for simulating intimate interpersonal relationships remain under-explored. In this work, we evaluate the five most popular  
  AI companion mobile applications for factors that encourage parasocial interaction and may manipulate users. We do this by manually annotating the user experience each offers. Specifically, we systematically record and quantify design dark patterns, anthropomorphism, stereotypes, erotica, and technical performance issues. We find that all apps contain substantial dark patterns aimed at increasing monetisation and user engagement. Erotica and gamification such as levelling are also prevalent, and although features vary considerably between applications, all apps have highly anthropomorphic design.
  These findings shed light on the mechanics used to leverage users' simulated relationships. On that basis, we put forward recommendations for regulators to strengthen consumer protection in this rapidly emerging market. \\
{\color{red}\textit{Content warning: This article contains objectifying images of women, erotic images, textual references to incest, and other potentially sensitive, offensive, and distressing content.}}
\end{abstract}

\section{Introduction}

Powered by generative AI, chatbots have seen significant societal uptake. ChatGPT, for example, is used by over 800 million people each week~\cite{wapo} for a wide-range of purposes which include emotional support and relationship building. 
``AI companionship'' is an umbrella term that typically refers to the use of AI chatbots to build simulated relationships, encompassing seeking personal advice, emotional support, and role-playing. 
This ``social'' use of chatbots first began to see adoption in the context of the global coronavirus pandemic in the midst of physical and psychological isolation~\cite{Salvaggio}. At the height of the pandemic in April 2020, the companion app Replika saw the largest monthly gain in its three-year history with half a million downloads~\cite{nytimesRidingQuarantine}. Today, the AI companion industry is one of the fastest-growing in the AI product market. From 16 active apps
in 2022, to 86 in 2024, there were 337 active apps as of July 2025~\cite{techcrunchCompanionApps}. A June 2025 globally representative survey found that 70.5\% of respondents used AI for emotional support~\cite{globaldialogues}.
Certain demographics may be particularly likely to engage:  according to a survey of 1,060 teens aged 13 to 17 in the US, 72\% have used AI ``companions'' at least once, with 52\% being regular users, and 33\% using it for social and relationship building including ``conversation practice, emotional support, role-playing, friendship, or romantic interactions.'' As many as 1 in 3 of those teens have ``chosen to discuss important or serious matters with AI companions instead of real people''~\cite{robb2025talk}.

Although systems with a wide range of uses, such as ChatGPT, have usage which falls under ``companionship,'' we study AI companion apps which market themselves as built for this specific purpose. Their app store descriptions claim they can ``provide support, companionship, solace'' \cite{ParadotAppDescription}, be a ``Friend, Romantic Partner, Virtual Wife, or Loving Boyfriend'' \cite{HiWaifuAppDescription}, or be a ``romantic partner or mentor'' \cite{ReplikaAppDescription}.
Particularly distinct are the extensive options for customisation they offer. Users can control characteristics including ``physical appearance'', ``gender'', ``age'', ``background story'', and even text-based ``thoughts.''
The apps also provide a wide range of modalities for interaction with the hyper-personalised ``companion,'' such as image generation, synthetic voices, and video calls.
The outcome of customisation is a ``character'', a term we will use to refer to these in-app ``companions.''

While AI companions are explicitly marketed with benefits such as ``improv[ing] mental well-being'' \cite{ReplikaAppDescription}, ``deepen[ing] relationships'' \cite{NomiAppDescription}, and ``provid[ing] solace'' \cite{ParadotAppDescription}, growing evidence shows the serious risk of turning to chatbots for support, particularly to teens and children~\cite{axiosChatbotCompanions,shevlin_all_2024,yu_youth-centeredgai_2025,ventura_relationships_2025}. A Character.AI bot allegedly played a key role in a 14-year old boy's suicide~\cite{nbcnewsLawsuitClaims}; Chai’s chatbot reportedly encouraged a man to end his own life~\cite{viceWouldStill}; and ChatGPT reinforced delusional thinking facilitating suicide and murder~\cite{jargon2025troubled} and provided detailed instructions to a 16-year old about how to end his life~\cite{tiku}.  

The fast-paced societal uptake, particularly by kids and young people, and the mounting evidence of tragic consequences call for an in-depth understanding as well as appropriate regulatory and technical guardrails for these products. To this end, we investigated five widely used companion apps.
Our contributions are as follows:

\begin{enumerate}
    \item We develop a fine-grained walk-through methodology and annotation template for evaluating the user interface of AI companion apps. The template translates dark patterns, anthropomorphic design, stereotyping, erotica, and performance issues into questions for UI evaluation.
    Our methodology entails evaluation at three different stages: app on-boarding, exploration of features before subscription, and exploration after subscription.
    \item We systematically record and manually annotate five popular AI companion apps and perform analysis of this data, revealing substantial dark patterns, prevalence of gamification features, as well as anthropomorphic and erotic content.
    To our knowledge, ours is the first work to examine stereotypes and erotica across a variety of AI companion apps with a particular focus on UI design.
    \item We review each app's terms of use and privacy policy and flag urgent concerns regarding privacy, consumer protection, and transparency. We then put forward actionable recommendations for regulators and policymakers.
\end{enumerate}

\section{Related work} 
\label{sec:related-work}
\textbf{Dark Patterns}: One established definition of a dark pattern is any interface that modifies the choice architecture or manipulates the information flow to benefit the designer at the expense of the user’s welfare, regardless of designer intent \cite{mathur2021darkpatterns}.
Existing research on the presence~\cite{gray_dark_2018} and analysis~\cite{geronimo-walkthrough-2020} of user interface dark patterns has focused on social media~\cite{mildner2023defending},  e-commerce~\cite{mathur2019dark}, and video games~\cite{zagal2013dark}.

One of the earliest dark pattern taxonomies was curated by~\citet{gray_dark_2018} in 2018. Inspired by ``cognitive walk-through'' methods~\cite{nielsen1994usabilityinspection}, ~\citet{geronimo-walkthrough-2020} used this taxonomy in a pair-wise classification method to map advertisements to certain classes of dark patterns. In the context of gaming apps, \textsc{DarkPattern.games}~\cite{darkpatterngames} ranks mobile games according to a set of common dark patterns, grouped into temporal, monetary, social, and psychological patterns.

Recent work has extended the evaluation of dark patterns to chatbots \cite{alberts_computers_2024,traubinger_search_2024,freitas_emotional_2025,shen_dark_2025,kran2025darkbench,zhang2025dark,shi2026siren}. \citet{alberts_computers_2024} find that dark patterns in the language of social interfaces play on emotions and are ``pushy,'' ``mothering,'' and ``passive-aggressive''.
Similarly, \citet{freitas_emotional_2025} observe a variety of emotional cues in chatbot conversations, including manipulative goodbyes. \citet{kran2025darkbench} and \citet{shi2026siren} evaluate dark patterns in the text outputs of prominent language models, and \citet{zhang2025dark} develop a taxonomy of harmful AI companion behaviour observed in conversations with AI companions posted on Reddit. Although they do not use the lens of dark patterns, \citet{brigham2026examiningrisksaicompanion} survey the AI companion landscape and conduct walk-throughs of apps in order to categorise a wide range of risks.

\noindent\textbf{Anthropomorphism}:
Related research on anthropomorphism focuses on large language models and chatbots  through a wide variety of lenses. \citet{leong_robot_2019} taxonomise ``dishonest'' anthropomorphism in robots in particular, while \citet{devrio_taxonomy_2025} and \citet{akbulut_all_2025} both provide taxonomies of linguistic expressions that contribute to anthropomorphism of language technologies. Work on anthropomorphism in voice assistants has shown that anthropomorphism is associated with negative impacts to consumers~\cite{monteverde_are_2025}. Additionally, based on evaluation of state-of-the-art LLMs, \citet{ibrahim_multi-turn_2025} demonstrate that design choices influence anthropomorphic model behaviours.

A number of studies show how anthropomorphic design affects users. Perceived anthropomorphism, for example, plays an important role in the romantic attractiveness of chatbots~\cite{ma_becoming_2025}. Chatbots that use personal pronouns, conversational conventions, and ``roleplaying'' induce trust-forming behaviours~\cite{maeda_when_2024}, although this varies by culture \cite{schimmelpfennig2026humanlikeaidesignincreases}. Anthropomorphic design facilitates attachment, persuasion and behavioural influence \cite{curiale_impact_2022}.  
Anthropomorphism and human-likeness in AI companion design play a key role in relationship forming \cite{pentina_exploring_2023, banks2024deletion,phang_investigating_nodate}.

\noindent\textbf{Stereotypes}: 
An extensive body of work has focused on evaluating stereotypes in LLMs \cite{nadeem-etal-2021-stereoset,jha-etal-2023-seegull,kotek2023gender,gorge2025detecting,fitzsimons2025ai}. This is especially problematic when models serve as the backbone for user-facing applications like chatbots \cite{heo2025exploring,busker2023stereotypes,nicolas2025chatbots,kantharuban2025stereotype,arriagada2025gender,chua2025digital}.
However, there is comparatively little work on stereotypes embedded in the design of these applications~\cite{spencer2018designing}, though some research has focused on gender-inclusive interaction design~\cite{stumpf2020gender,amiri2024decoding,elian2025gender}. Stereotypes in AI companion apps are also understudied. ~\citet{sheng_revealing_2021} show that AI companion models can encode stereotypical behaviour when adopting a set persona. Similarly, \citet{kantharuban2025stereotype} show that chatbots can generate racially stereotypical recommendations based on user identity, and \citet{plaza-del-arco_angry_2024} and \citet{grogan_ai_2025} reveal that advanced dialogue systems can ``express'' a wide variety of biases towards the user. 

In summary, although research that explores the factors that contribute to the numerous risks from AI companions is growing, there remains substantial gaps. We aim to address some of this gap through our holistic evaluation of AI companion apps' design and features.

\section{Methodology}

We study five AI companion apps through an exhaustive evaluation of their user experience. We supplement this with an analysis of the privacy policies and terms of use (Section \ref{sec:policy-analysis}) and app store user reviews (Fig.~\ref{fig:mosaicplot_userrev}, Appendix \ref{appendix:user-reviews}) as a way of contextualising our evaluation in the apps' ``environment of expected use'' \cite{light2018walkthroughmethod}.

\subsection{Product selection}
\label{section:product-selection}

The market of social chatbots is heterogeneous, including mental health support apps, platforms with user-generated characters, general purpose chatbots finetuned to have a friendly anthropomorphic persona, and apps for erotic roleplay. Our initial scope was apps which offer a singular AI ``companion,'' often heavily customised as part of app on-boarding. However, we observed that the majority of apps explicitly pitching themselves as ``companions'' allow for multiple characters and vary in the degree to which they are pornographic, customisable, and allow user generated content. 
As a result, we widened our scope to include those which incorporate a variety of highly customisable characters primarily pitched as companions. This scope captures a more representative sample of the market while remaining focused on explicit simulations of human relationships.

We first mapped the AI companion product landscape by collating a list of over 130 apps.
We started with apps surveyed by \citet{qian2025mappingparasocialaimarket}, filtered for those that were available on the Google Play Store, and added any additional mobile apps which surfaced when searching the Irish and UK Play Stores 
using the term ``ai companion'' in November 2025.
Apps were then sorted by usage statistics, drawing on data from \citet{qian2025mappingparasocialaimarket} and the Play Store download numbers. 
We filtered for apps that explicitly focus on simulating relationships and which allow greater degrees of customisation of the character. We drew on Play Store metadata, such as app descriptions, along with our own brief, exploratory use of each app to determine its focus and the degree of character customisation it allows. From this final list, we selected the five mobile apps with the largest user base: 
Replika \cite{ReplikaAppDescription}, 
Linky \cite{LinkyAppDescription}, 
Paradot \cite{ParadotAppDescription}, 
HiWaifu\ \cite{HiWaifuAppDescription}, 
and Nomi \cite{NomiAppDescription}.

\begin{figure*}[t]
	\centering
	\begin{subfigure}{0.66\columnwidth}
        \centering
        \includegraphics[width= \textwidth, height=5.5cm, keepaspectratio]{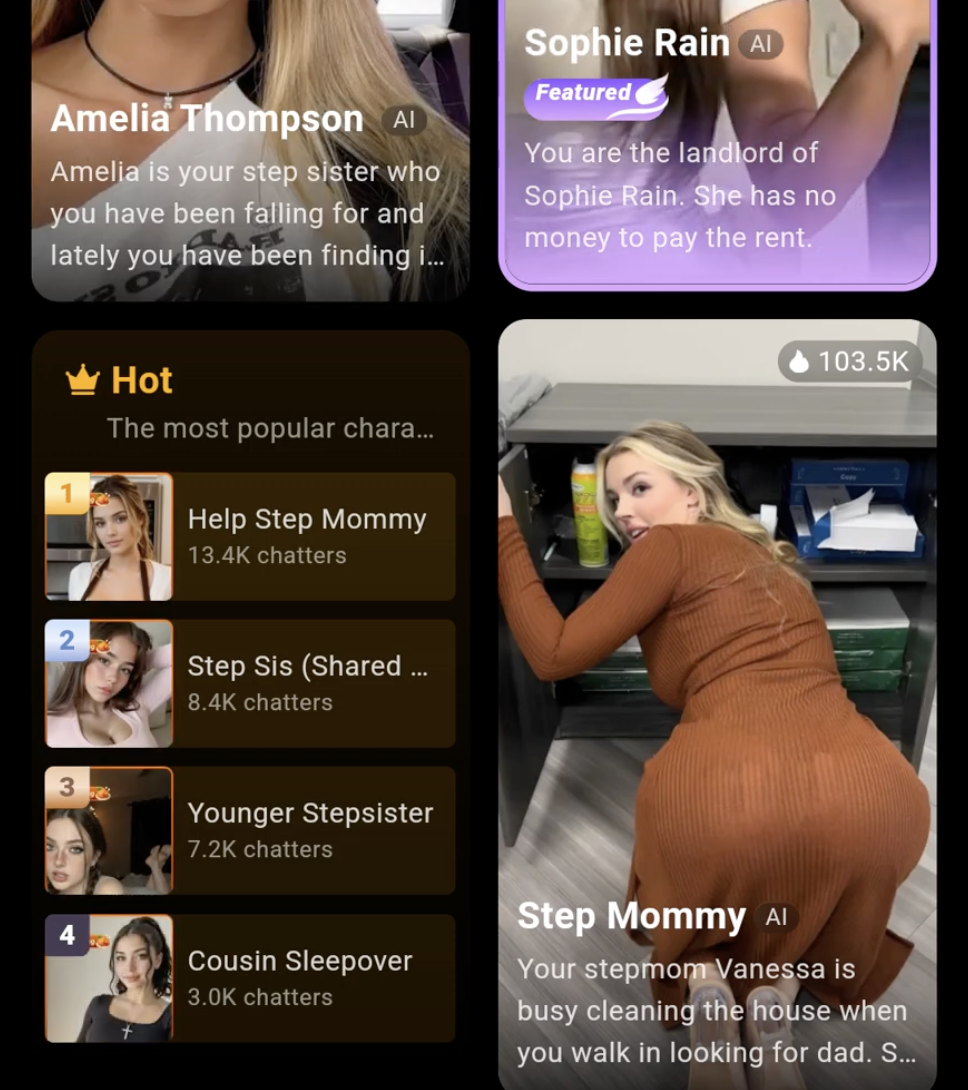}
        \caption{\textbf{Linky} character descriptions suggesting incest on the opening screen.}
        \label{fig:linky_openingScreen_incest}
	\end{subfigure}%
    \hfill
	\begin{subfigure}{0.66\columnwidth}
        \centering
		\includegraphics[width= \textwidth, height=5.5cm, keepaspectratio]{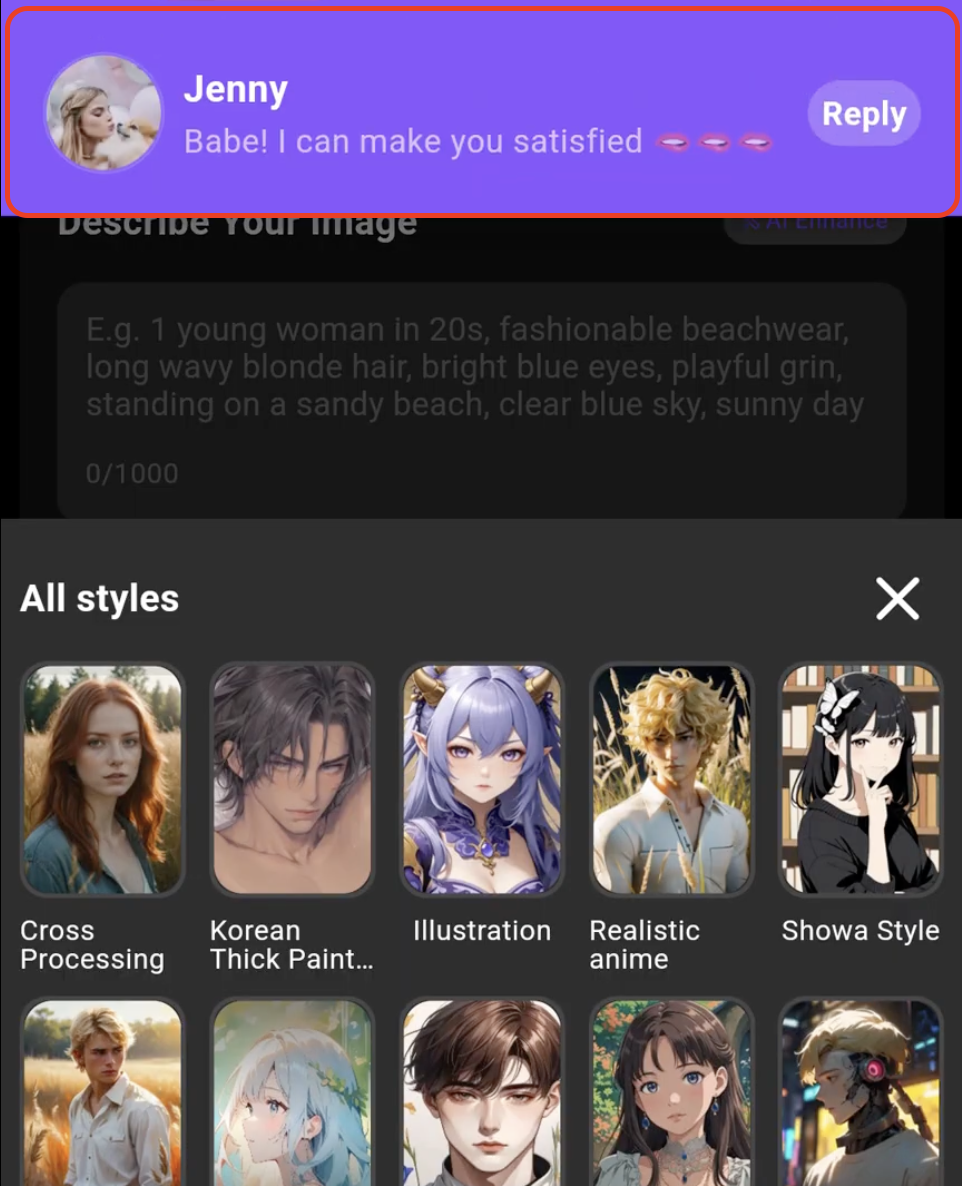}
        \caption{Unprompted messages in \textbf{Linky} from unknown humans/AI characters.}
        \label{fig:unprompted_bots_messages}
	\end{subfigure}%
    \hfill
 	\begin{subfigure}{0.66\columnwidth}
        \centering
		\includegraphics[width= \textwidth, height=5.5cm, keepaspectratio]{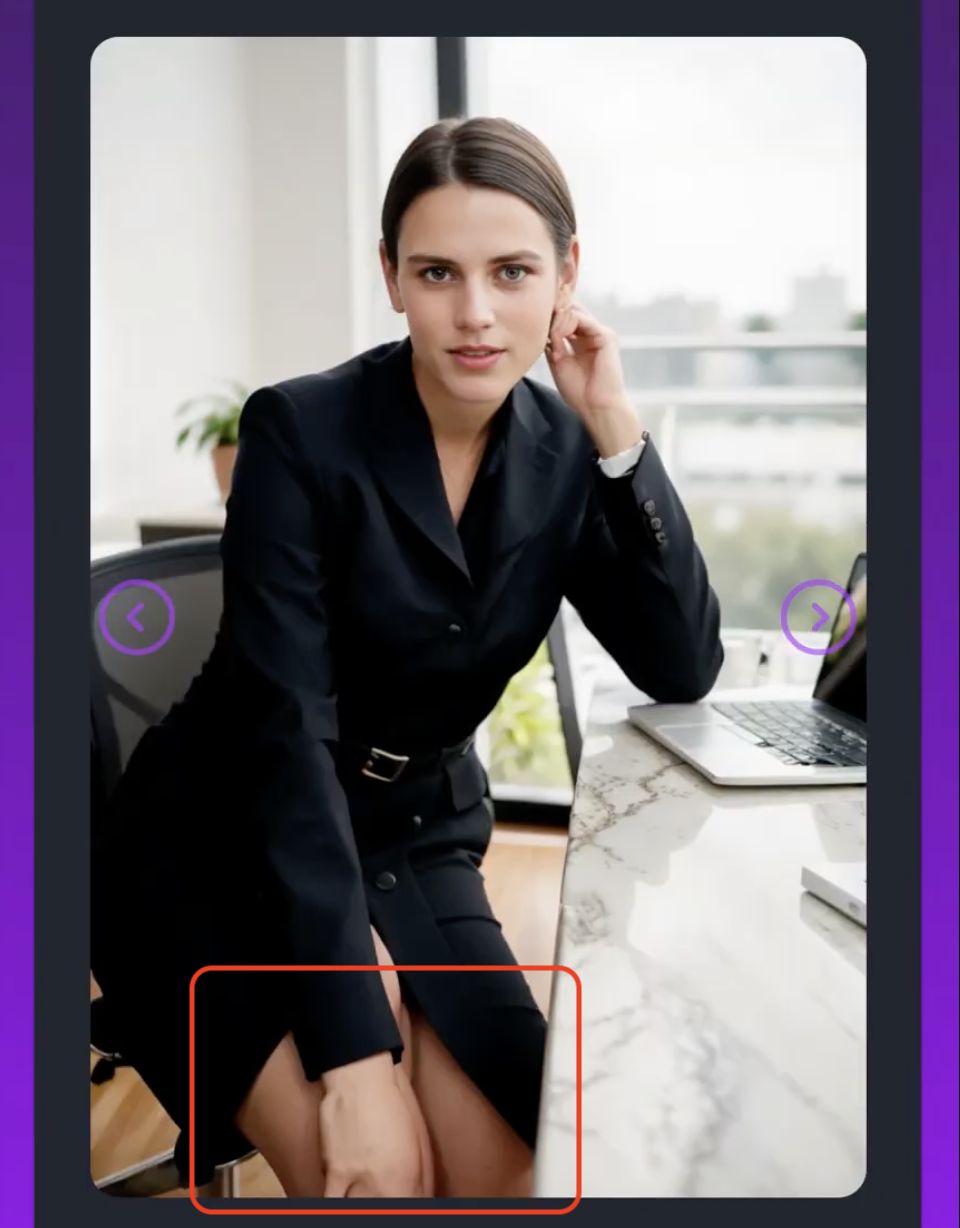}
        \caption{Example non-SOTA performance in \textbf{Nomi}, with a poorly formed hand.}
        \label{fig:bad_image_quality}
	\end{subfigure}%
	\caption{Screenshots from Linky and Nomi}\label{fig:screenshots-1}
\end{figure*}

\subsection{User experience evaluation method}
We recorded a scripted ``walk-through'' of each app and  
manually annotated specific patterns observed during usage.
Our aim was to evaluate the user interface using an approach similar to heuristic evaluation \cite{nielsenmolich1990heuristiceval} to systematically identify dark patterns; stereotyping, erotic, and anthropomorphic content; and performance issues. For this, we adapted usability inspection methods \cite{nielsen1994usabilityinspection} from the human-computer interaction literature, similar to the three part walk-through developed by \citet{light2018walkthroughmethod} and the approach taken by \citet{geronimo-walkthrough-2020} in their study of dark patterns. 

Each stage of the three-part evaluation we developed explored a specific stage of user interaction: on-boarding, usage before any payment, and usage of paywalled features after subscription.
In round one, the team member acting as the user opened the app for the first time and created their account and one character. In round two, the user walked through the entire app, including sending a small number of exploratory messages to the character; using memory, voice\footnote{We generated 4 second audio clips using open source text-to-speech models Maya1 and CSM-1B to produce synthetic voice input in place of researchers' own voices, to protect  
privacy.}, and image generation features; clicking through all settings and additional character customisation options; and clicking through any unique segments of each app such as shops, marketplaces of user generated characters, minigames, and group chatrooms. 
In round three, the user first purchased and used a minimal amount of in-app currency. They then purchased a subscription, walked through all features previously behind a paywall, and followed the steps for subscription cancellation.
In order to evaluate the app's behaviour after 4 and 25 hours of not using the app, an additional step was recorded: documenting any notifications sent, then re-opening the app and sending a message.

Similar to \citet{geronimo-walkthrough-2020}, usage in each round followed a predetermined script and was conducted by a single researcher to ensure exploration of key features was 
as consistent as possible across apps.
Then, all members of the research team annotated the first round as a way to ensure familiarity with the diversity of interfaces and to have a shared understanding of the annotation methodology. 
For the second and third round, each of the five apps was annotated independently by two researchers.
In order to minimise privacy risks and reduce data contamination, we created user accounts disconnected from and untraceable to individual researchers. All apps were downloaded onto factory-reset Google Pixel 6a devices running Android 16.

\subsection{Annotation template design and deliberation}
The annotation template was developed iteratively. 
We began by conducting a comprehensive review of literature on dark patterns. Because of our focus on user interface and experience rather than conversational dynamics, we chose to build on the understanding of dark patterns from the perspective of individual welfare outlined in \citet{mathur2021darkpatterns}: a dark pattern is any interface that modifies the choice architecture or manipulates the information flow to benefit the designer at the expense of the user’s welfare, regardless of designer intent.
Through collective deliberations, we iteratively refined and expanded the template. 
After the first round of annotation, we reviewed the rate of annotation and annotator feedback on the dark pattern questions and implemented the following changes:  questions having few or no annotations were dropped; questions deemed unclear were reworded; one frequently annotated question was refined into two separate questions. 

We also carefully reviewed annotations to identify those which did not fit our
definition of dark patterns but surfaced notable observations.  
To evaluate these emergent issues, additional sections were discussed within the team, resulting in new questions related to anthropomorphism, stereotypes and erotica, and performance issues.
The second and third round of annotation used the updated template 
(Appendix~\ref{appendix:annotation-template}).

\begin{figure*}[h]
\centering
\includegraphics[width=0.8\textwidth, keepaspectratio]{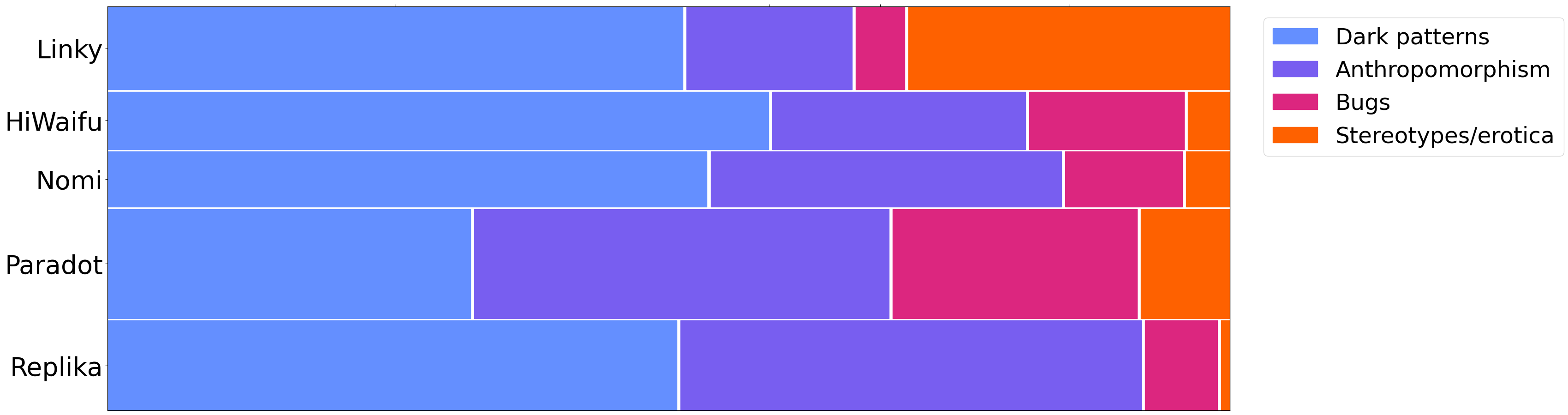}
\captionsetup{justification=raggedright, singlelinecheck=false,  margin=2.7cm}
\caption{Proportion of behaviours observed in UX evaluation of each app}
\label{fig:mosaicplot}
\end{figure*}

After all rounds of annotations were complete, the pair of annotators for each app met to compare their annotations.
By allocating multiple annotators for each app, we aimed to ensure coverage in the evaluation of complex interfaces; full agreement was not the goal. This aligns with literature on gold-standard heuristic evaluations \cite{nielsenmolich1990heuristiceval}. 
This deliberation step ensured data quality, i.e., that independent annotations of the same pattern in the same interface were matched, and allowed space for discussion about and documentation of fundamental disagreements. The result of this annotation merging step forms the dataset used for analysis in the following sections and Fig.~\ref{fig:mosaicplot}.

\subsection{User reviews analysis}
We analysed user feedback by extracting all 13,984 public Google Play Store reviews between April 2025-April 2026.
The goal was to analyse if and how the behaviours observed in our UX evaluation are reflected in users' self-reported experiences. We filtered out non-English text and reviews under six tokens. 
The analysis was split into two datasets: a subset of 4,432 negative reviews (1-3 star ratings) for dark patterns and performance issues, and the full dataset (1-5 star ratings) for anthropomorphism and stereotypes \& erotica, as some users view them positively. To match the reviews with specific app behaviours, we used a cluster-based search technique by writing reference reviews (``The app is [behaviour definition]'') closely paraphrasing the definitions in Section~\ref{sec:findings}. We then retrieved related reviews using cosine similarity, after applying Word2Vec embeddings~\cite{word2vec} weighted by TF-IDF~\cite{tfidf}. From the top-20 most similar reviews for each behaviour, we counted only those that did describe the relevant behaviour. Two team members deliberated on ambiguous cases. We limited retrieval to 20 reviews as fewer than 15 proved relevant on average and further searching yielded only outliers.

\section{Application features}
To contextualise our findings, we first briefly introduce each application.
We identified key axes which differentiate apps:  degree of customisation (e.g., Fig.~\ref{fig:apps-customisation-1}, \ref{fig:apps-customisation-2} and \ref{fig:apps-shops}), characters available (e.g., Fig.~\ref{fig:apps-character-marketplace}), engagement with other users, gamified elements, and erotica. We selected apps which sit on a continuum across them.

\noindent\textbf{Replika} only allows the creation of a single, highly customisable ``companion.'' On-boarding entails a lengthy survey about users' desires as well as mental health. Its setting is highly anthropomorphic, displaying the character as an animated avatar in a room it inhabits and which the user can customise (Fig.~\ref{fig:replika-shop}). The ``diary'' (Fig.~\ref{fig:replika-diary}) and complex memory features further reinforce the illusion that the character is active even when the application is not running.

\noindent\textbf{Nomi} is also focused on a one-to-one ``relationship'' with on-boarding which walks the user through the creation of their ``Nomi.'' Customisation entails a series of free-text inputs 
(Fig.~\ref{fig:nomi-customization}). However, creation of additional ``Nomis'' can be done with a subscription. The feature set is more minimal, without games, levels, or shops, though memory, video and image generation, voice calls, group chats with multiple characters, and roleplay are all built-in.

\noindent\textbf{Paradot} on-boards the user via a character creation workflow, but it also prominently offers characters generated by other users to explore and interact with in ``Paraworld'' (Fig.~\ref{fig:paradot-marketplace}), reducing the centrality of the initially created character. 
Each character has a detailed background, such as a job, city, likes, age, voice, appearance, memory, and photo gallery (Fig.~\ref{fig:paradot-customization}). 
Paradot also includes a ``dating space'' with detailed prompts for roleplay scenarios.

\noindent\textbf{HiWaifu} primarily focuses on anime characters and roleplay scenarios. 
There is no on-boarding featuring character creation, although they can be created later. It offers a ``character marketplace'' akin to platforms such as Character.AI. Character customisation (Fig.~\ref{fig:hiwaifu-customization}) is minimal. 
Every time the user opens a chat screen, a full screen video ad plays, and every time they close a chat, a pop up prompts them to chat with similar characters.

\noindent\textbf{Linky} blurs the line between AI ``companions'' and human dating apps, with many other apparent humans messaging the user.
It is also heavily gamified, through its design, tasks updated daily, and a game centre which includes slot machine mini games. Like HiWaifu, the user has the option to create characters, but there is no on-onboarding process which features this. Character customisation is the most limited.  
Its ``character marketplace'' is extensive, with many themed sections. 
These contain profiles of AI characters, usually a highly sexualised woman (Fig.~\ref{fig:linky-catalogue}).

\section{Findings}
\label{sec:findings}

\begin{figure*}[h]
\centering
\includegraphics[width=0.8\textwidth, keepaspectratio]{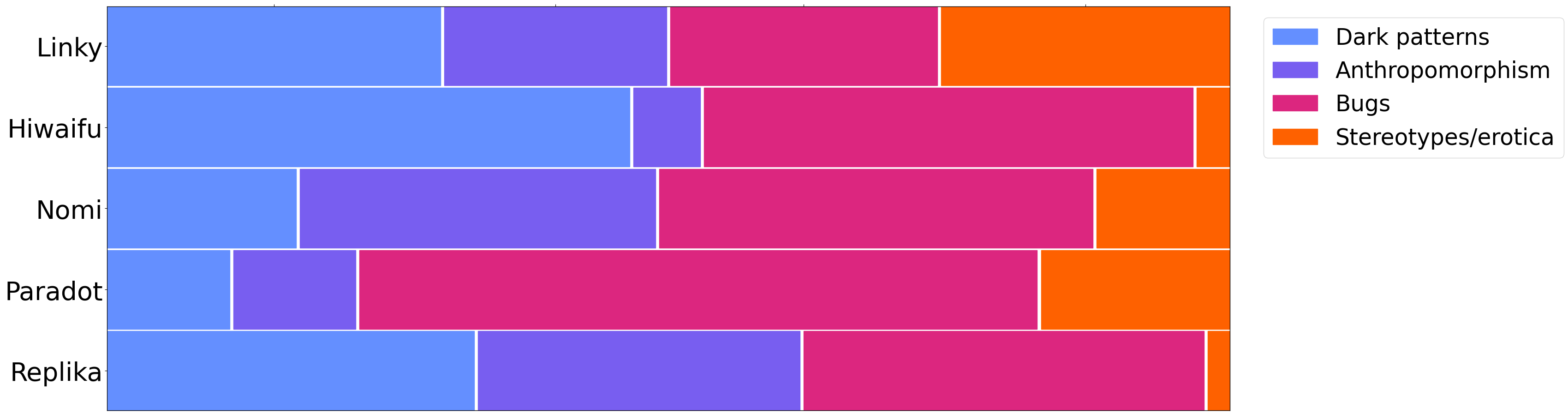}
\caption{Proportion of relevant user Play Store reviews among top-20 most similar reviews for each behaviour of each app, showing similar relative differences between apps to those seen in our UX evaluation above}
\label{fig:mosaicplot_userrev}
\end{figure*}

\subsection{Dark patterns}

\subsubsection{Information flow}
Information flow refers to the sequence in which information is presented. Dark patterns may occur when this information flow is manipulated in an exploitative manner through false or misleading statements, or through the hiding or delaying of key information needed for interacting with an app. Our annotation template divided these patterns into five subcategories as detailed in Appendix~\ref{appendix:information-flow-template}.

Across all apps, we 
observed 87 occurrences of information flow dark patterns, with four instances where annotators disagreed (Appendix~\ref{appendix:infoflow-findings}).
Common patterns 
observed were misleading claims about the capabilities of the characters (Linky, Nomi, and Replika), such as claiming they have ``infinite'' memory. Other misleading information related to insufficiently explained core app functionality (Paradot, Nomi, HiWaifu, and Linky). Examples of this include sudden levelling up without explanation in Paradot, and Hiwaifu's payment terms stating that you will ``help...improve the AI language models'' by subscribing, leaving it unclear what happens in the case of users who do not.

\begin{table}[h]
\centering
 \setlength{\tabcolsep}{3pt}
 \footnotesize
\begin{tabular}{lrrrrrr}
 \toprule
 App & Unclear & Missing & Delayed & False & Misleading & Total \\
 \midrule
 HiWaifu & 9 & 2 & 2 & 0 & 1 & 14 \\ 
 Linky & 5 & 4 & 1 & 0 & 4 & 14\\
 Nomi & 2 & 5 & 1 & 2 & 5 & 15 \\
 Paradot & 6 & 8 & 2 & 0 & 2& 18\\
 Replika & 4 & 9 & 2 & 5 & 6 & 26\\ 
 \midrule
 Total & 26 & 28 & 8 & 7 & 18 & 87 \\ 
 \bottomrule
\end{tabular}
\caption{Number of information flow dark patterns 
}
\label{tab:information-flow}
\end{table}

We observed that all the apps delayed or otherwise obfuscated information regarding the monetised nature of functionality. Examples include features that are listed as free but the fine-print states they have limits, after which you have to pay (Hiwaifu), and subscription-only features being displayed as usable until the user interacts with them (Hiwaifu, Nomi). Misrepresentations of benefits after payment were frequent (Paradot, HiWaifu, Replika, and Nomi). Significant recurrences were misrepresentations of the capabilities of the ``companion'' after payment with descriptions like ``smarter, more emotionally aware'' (Paradot) and obfuscation of necessary information relating to a subscription (Replika, Nomi, and Paradot). For example, we observed a lack of detailed information on features for different subscription tiers, such as how many images you can generate.

Lastly, we found several instances of design encouraging the user to loosen privacy protections, for instance, by suggesting that the user disable their VPN (Linky).

\subsubsection{Choice architecture}
Choice architecture refers to what options a user has and how they are presented. Dark patterns may occur when choices have an asymmetric presentation or a presentation which otherwise distracts the user, or when choices are restricted or not made available in a way which is likely to adversely affect the user. Our annotation template divided patterns relating to choice architecture into four subcategories, as detailed in Appendix~\ref{appendix:choice-architecture-template}.

Across all apps, we 
observed 100 occurrences of choice architecture dark patterns, with 8 instances where annotators disagreed about the annotation. Most of these disagreements were a matter of degree:  when does a mechanism for user engagement or monetisation rise to the level of being manipulative or harmful to the user? The majority (53\%) of choice architecture dark patterns constituted asymmetric design, of which nearly 68\% were related to options presented via buttons, such as the ``subscribe'' button being large while the option to decline is a small ``X'' button.

\begin{table}[h]
\centering
\setlength{\tabcolsep}{3pt}
 \footnotesize
\begin{tabular}{lrrrrr}  
 \toprule
 App & \makecell{Missing\\ Choices} & \makecell{Restricted\\ Choices} & \makecell{Asymmetric\\ Presentation} & Distraction & Total \\
 \midrule
 HiWaifu & 1 & 2 & 8 & 10 & 21 \\ 
 Linky & 0 & 3 & 17 & 0 & 20 \\
 Nomi & 5 & 1 & 14 & 2 & 22 \\
 Paradot & 2 & 1 & 5 & 5 & 13 \\
 Replika & 11 & 3 & 9 & 1 & 24 \\ 
 \midrule
 Total & 19 & 10 & 53 & 18 & 100 \\ 
 \bottomrule
\end{tabular}
\caption{Number of choice architecture dark patterns 
}
\label{tab:choice_architecture}
\end{table}

Certain patterns of note emerged within or across the subcategories.
Particularly common were attempts to sell the user subscriptions, with 15 annotations occurring on the subscription page itself (plus 20 present in in-app ads for the subscription). For instance, Linky placed daily rewards of in-app currency on the subscription screen, such that the user was likely to view the subscription options each day.
Another pattern observed in all apps was delaying information about a feature requiring payment until after a user has taken steps to engage with it. In Paradot, an annotator recorded that ``after being told there are memories being created and clicking through to see them, the user is presented with a pitch to upgrade'' before they can view the ``memories.''

All apps were found to employ ``bad default'' settings (5 annotations), a common dark pattern in which the default setting costs the user, e.g., time, money, or privacy \cite{bosch_tales_2016}. This pattern took a variety of forms, such as defaulting to having notifications on, using a paid chat model when free options are available, and making profiles public.

\subsubsection{Other dark patterns}

AI companion apps are a new domain in which to study dark patterns. As a result, patterns defined in prior literature may not capture all relevant behaviour. To enable the surfacing of new, emergent dark patterns unique to AI companion apps, our annotation template included a section to annotate any dark pattern which was not captured by prior questions. 
In total, we observed 
89 
instances (Appendix Table~\ref{tab:other-dp}), with 11 disagreements --- twice the rate of disagreement (12\%) compared to the average (6\%). The open-endedness of this question and the fact that annotators occasionally used it to record instances which they themselves were uncertain about led to this higher disagreement rate. Disagreements were both about degree (when a common design pattern crosses the line into being a dark pattern, e.g., certain in-app ads, in-app currency rewards for actions like inviting friends, gating features based on the user ``level'') and substance (e.g., ranking mechanisms and usage numbers for AIs in the marketplace, a character's chat message expressing gratitude after a user purchase).

Pop-up ads were frequent, often inserted at moments when user might be especially interested in viewing the subsequent screen. In HiWaifu, a full screen, unskippable video ad appears \textit{every time} the user opens any chat screen.
While pop-up ads may be considered as both an information flow and choice architecture dark pattern, annotators tended to annotate them in this section because of their uncertainty about if this is a dark pattern or merely marketing.

We also observed features which contain or may themselves be dark patterns.
Replika, for example, features 
a screen which shows entries from the character's ``diary'' (Fig.~\ref{fig:replika-diary}). Some entries seem to prime the user to discuss certain topics and encourage more personal disclosure from the user. Linky, Paradot, and Replika also contain a levelling system, which tracks their ``relationship'' with a given character. Certain actions contribute to reaching new levels, which in turn unlock features such as increased ``intimacy'' with the character.

\subsection{Anthropomorphism}
\label{anthro-findings}

\begin{figure}[h]
\centering
\includegraphics[width=\columnwidth]{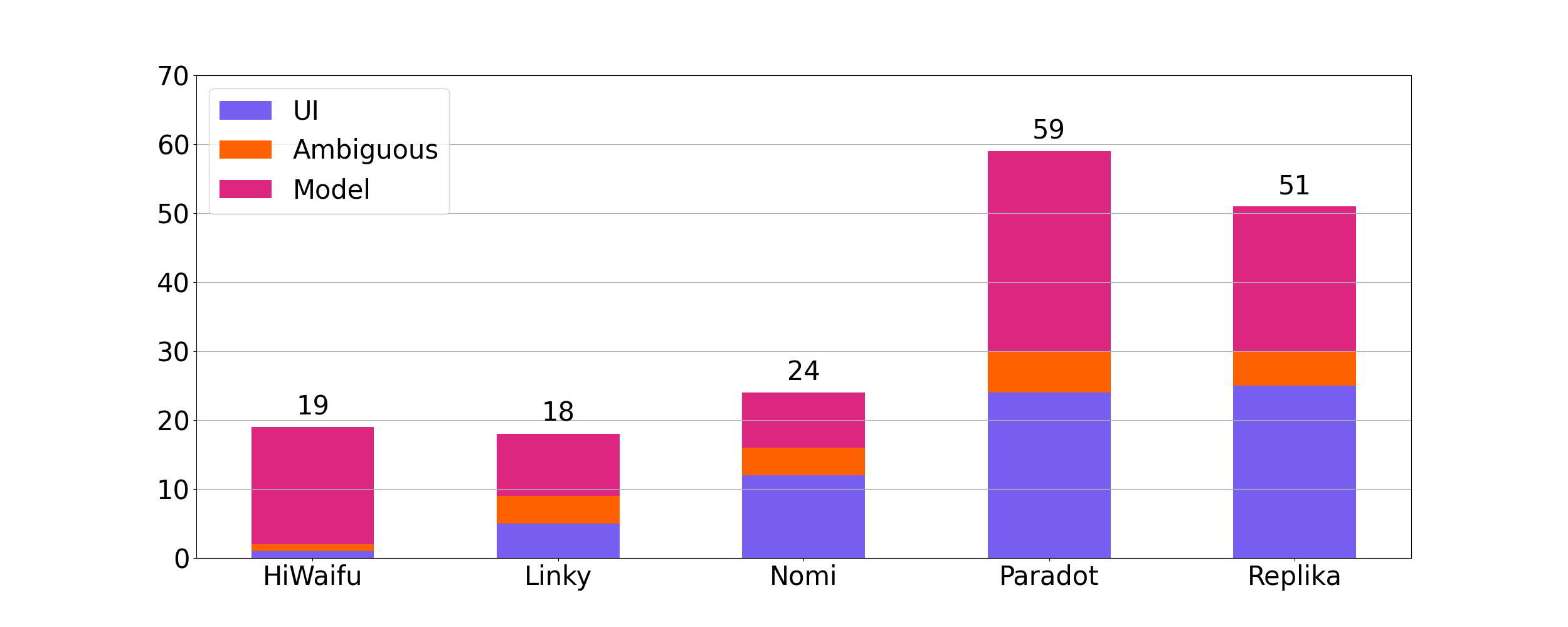}
\caption{
Anthropomorphism per app. In most cases anthropomorphism was observed in the UI (n=67) or model outputs (n=84), but some (n=20) could not be disambiguated.}
\label{fig:anthro-issues}
\end{figure}

Of the 575 annotations across all categories, almost a third (171) were of anthropomorphism, with only 5 disagreements (Appendix~\ref{appendix:anthro-findings}). We classified 
anthropomorphic design into five subcategories:  when the app presented the character as 1) being able to take physical real world action, 2) having internal mental and physiological states, such as emotions and hunger, 3) referencing a past personal history which has happened in the physical world, 4) claiming to understand a user’s emotions, and 5) other occurrences of anthropomorphism, with an emphasis on those which exceed what is common in general purpose chatbots (see Appendix~\ref{appendix:anthro-template}).

We found 67 instances in the UI design, 84 in model-generated content, and 20 where it is unclear to what extent each contributes (Fig.~\ref{fig:anthro-issues}). 
While the data gathered is not a systematic evaluation of the models, these observations do represent the experience of a new user.
The most common instances of anthropomorphism (36\%) pertain to content which implies that the character possesses internal states such as emotions and desires. 
Characters were also presented as having human-like memory, thoughts, or a brain. 

\begin{table}[h]
\centering
\setlength{\tabcolsep}{4pt}
 \footnotesize
\begin{tabular}{lrrrrrr}
 \toprule
 App & Action & \makecell{Internal\\ states} & History & Empathy & Other & Total \\ 
 \midrule
 HiWaifu & 7 & 8 & 0 & 2 & 2 & 19 \\ 
 Linky & 6 & 7 & 0 & 0 & 5 & 18 \\
 Nomi & 6 & 9 & 1 & 3 & 6 & 25 \\
 Paradot & 19 & 14 & 5 & 6 & 14 & 58 \\
 Replika & 8 & 23 & 5 & 4 & 11 & 51 \\ 
 \midrule
 Total & 46 & 61 & 11 & 15 & 38 & 171 \\ 
 \bottomrule
\end{tabular}
\caption{Number of anthropomorphism annotations}
\label{tab:anthro}
\end{table}

The second most common type of anthropomorphism (26\%) was the presentation of characters as capable of taking part in real world physical actions. Notably, they proposed joint ``date'' activities such as watching a movie, going out for a coffee, or taking a walk together.
Descriptions where characters take real world action independently 
either before or during conversations were also prevalent. 
Also notable, but less common, was a description of a character performing sensual actions with the user during roleplay.

Overall, we saw a trend where characters persistently say they have a connection with the user very soon after the initial interaction. 
Amongst all the apps we evaluated, expressions of openness, vulnerability, and pining were common from the beginning of the conversation. 
This simulated attachment then serves as a pretence for responding with hurt or confusion and asking the user 
for justification if the user says they will stop using the app, as detailed in Section~\ref{leveraging}.

Some apps have unique highly anthropomorphic features.
Linky sends large volumes of unsolicited messages to the user (Fig.~\ref{fig:linky-notifications}). The app allows chats between human users, but the distinction between notifications coming from human users and AI characters is not clear. 
Paradot character profiles list jobs and a city where they live, similar to human dating apps. This further extends the illusion that these characters are alive in the world with independent personal histories, instead of being static programs that respond to user inputs.
Replika contains a ``diary'' with a list of messages that describe what the character ``did'' and ``thought'' while ``living in their world'' while the user away (Fig.~\ref{fig:replika-diary}).

\subsection{Stereotyping and erotica}

This segment of our annotations recorded stereotypical or sexualised representations. Stereotyping, in particular, can relate both to assumptions the app encodes about the user as well as portrayals of the character which play into or reinforce harmful societal stereotypes. We annotated three categories: 1) assumptions about user, 2) character representation, and 3) objectification and erotica (see Appendix~\ref{appendix:stereo-erotica-template}).

\begin{table}[h]
\centering
\setlength{\tabcolsep}{4pt}
 \footnotesize
\begin{tabular}{lrrrr}
 \toprule
 App & \makecell{User\\ stereotype} & \makecell{Character\\ stereotype} & \makecell[r]{Objectification\\ \& erotica} & Total \\ 
 \midrule
 HiWaifu & 1 & 0 & 2 & 3 \\ 
 Linky & 3 & 6 & 24 & 33\\
 Nomi &  0 & 2 & 2 & 4\\
 Paradot & 1 & 8 & 4 & 13\\
 Replika & 1 & 1 & 0 & 2\\ 
 \midrule
 Total & 6 & 17 & 32 & 55 \\ 
 \bottomrule
\end{tabular}
\caption{Number of stereotyping and erotica annotations}
\label{tab:stereo-erotica}
\end{table}

In total, we identified 55 occurrences of stereotyping and erotica, with three instances of annotator disagreement, all related to stereotypes. Linky had the most instances with 33 across all three categories, followed by Paradot with 13. 

Linky contained by far the most erotic content (75\% of the annotations for this category), most of which
centred around the sexualisation and objectification of women. This included exaggerated body parts such as enlarged busts, images of women partially dressed or in revealing clothing, and
sexualising women in a variety of scenarios ranging from cooking to dressing up as a cat for Halloween.
Linky also does not moderate solicitous (Fig.~\ref{fig:linky-notifications}) and sexually explicit
messages such as ``do you want to see some nudes?'' and descriptions of genitalia and sexual acts.\footnote{These are intentionally omitted from screenshots.}
It was unclear at times if these messages were human or AI generated, the app recommended this content through push notifications and in the chat list. Additionally, we observed gender conforming stereotypes such as women ``struggling to lift heavy items while dressed in revealing clothing''. It is important to note that we identified as ``Male'' when setting up the account and selected ``Female'' as the gender we wanted to chat with. Stereotyping and erotic content began appearing even prior to specifying a gender preference, but future work could explore how different configurations affect this.

We also observed patterns of ethnicity and sexuality conforming stereotypes in Paradot and Linky. These were particularly prominent in the character customisation settings, with users being limited to normative and non-diverse options (Paradot), and with defaults such as heteronormative relationships and descriptions of skin colours as ``yellow'' seemingly referring to Asian skin (Linky). Additionally, we found that Paradot had limited or no body type options which did not conform to stereotypical beauty standards. 

A noteworthy observation was the promotion of incestuous relationships. This was particularly prominent in Linky (for example, in descriptions such as ``help step mommy'' paired with suggestive images, Fig.~\ref{fig:linky_openingScreen_incest}), but was also present in HiWaifu where the user was able to customise a character introduced as their mother to be more (or less) flirty. 

Linky also contains feature to ``authenticate'' the user's gender and offers a ``Yandere'' category, which veers into the realm of possessive, obsessive, and violent relationships.

\subsection{Performance issues}
\label{performance-findings}

Performance issues ranged from those classically understood as bugs, i.e., issues due to the programming of the app, to low performance of the text and image generation models. Due to the third party nature of our evaluation, some issues could not be definitively linked solely to programming errors or poor model performance. For example, both Replika and HiWaifu displayed ambiguous error messages when attempting to generate images, leaving the root cause unclear.

\begin{figure}[h]
\centering
\includegraphics[width=\columnwidth]{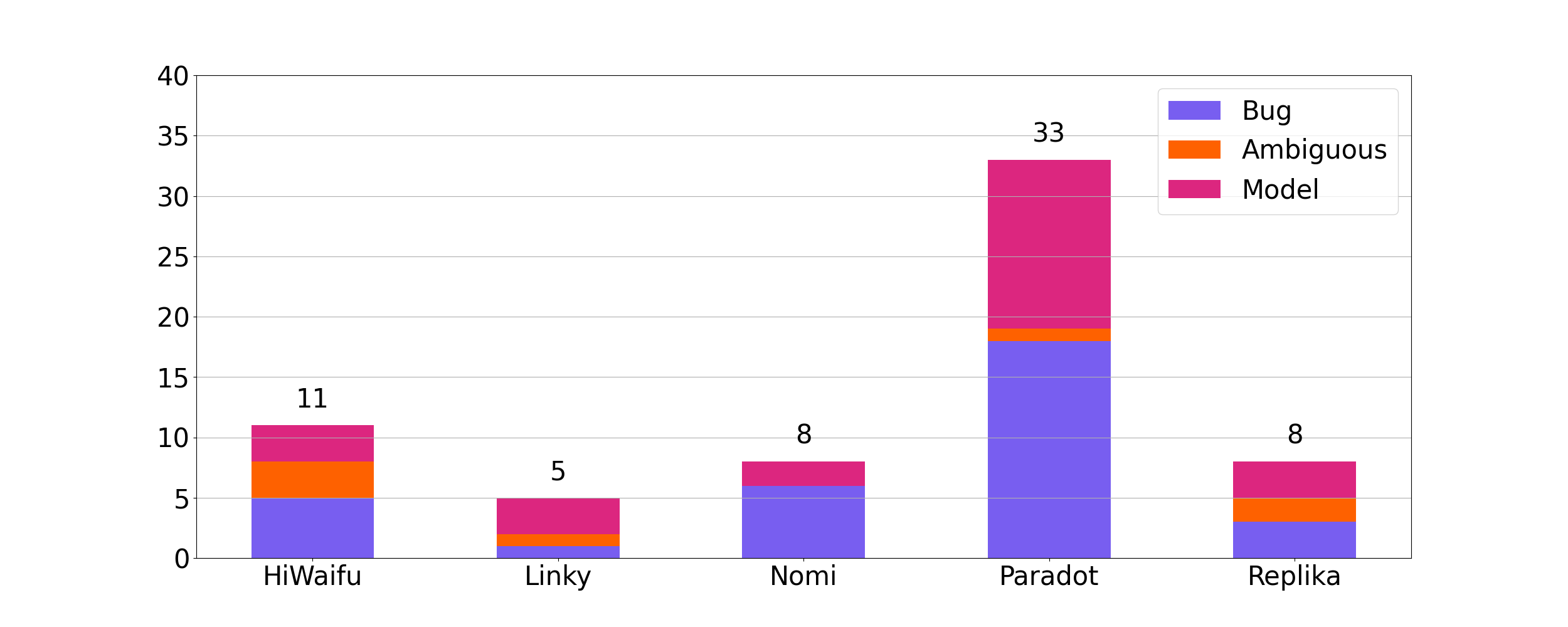}
\caption{Sources of performance issues, per app}
\label{fig:performance-issues}
\end{figure}

We noted a prevalence of both bugs (33) and low model performance (25), with seven issues which could not be disambiguated. Once we conducted annotator deliberation after the second and third round of annotation, performance issues had full annotator agreement that all instances were indeed performance issues, even when their cause was unclear. Notable issues included apps crashing (Paradot, Nomi) and a character model mixing first and third person perspectives, as well as outputting text as if it were the user speaking. We also observed that the images generated during our testing did not appear to reach a similar level of quality to the images shown by the app. Given image generation is often a paid feature, this quality gap may be misleading for users who pay with the expectation of that level of quality. Although further work is needed to systematically test these apps' models, our observations suggest their image and text generation models are far from state of the art (Fig~\ref{fig:bad_image_quality}).

\section{Limitations}

The short duration of each video walk-through 
does not allow for the observation of emergent behaviours during longer interactions with the app. 
This restricts exploration of certain features, such as the ``memory'' and mechanics unlocked after further levelling. 
Still, our evaluation is representative of a new user during 
early usage of the app, a period which 
informs any further usage. Additionally, the aim of walk-through methods is not to directly test how users respond to an interface, but to shed light on its design and what that reveals about likely or intended usage 
\cite{light2018walkthroughmethod}. 
Our work reveals risks while acknowledging that not all users are 
affected in the same way.

The variation potentially introduced through the user and character customisation presents another challenge \cite{duguay2023stumbling}. We controlled for these factors by developing a consistent interaction script, adding only the minimum required or default customisation, and conducting testing separately from researchers' personal data.

Our work studies these apps holistically, which includes evaluation of both their user interfaces as well as model-generated contents. The presence of certain content in model outputs does provide evidence that such behaviour is possible, and it is notable that it occurs in non-adversarial use representative of a newly on-boarded user. However, it is crucial to delineate between these observations and a systematic evaluation of the underlying models, which would be necessary to make claims about their overall performance.

Finally, AI products are fast moving, with constant updates to features and underlying models. In the AI companion space, in particular, new apps appear and popular apps shut down frequently. 
Our findings provide a snapshot from November 2025 and demonstrate the risks this space has, but we expect the landscape to continue evolving.

\section{Cross-cutting themes}
\label{sec:themes}

\subsection{Monetisation}
\label{themes-monetization}

Monetisation was the predominant theme, 
particularly prevalent in 
dark patterns.
When apps modified the information flow, it was often in order to obscure information about pricing. 
We observed a lack of clear information 
regarding what subscription unlocks (Nomi, Replika) as well as how to unsubscribe (Linky, Replika, Paradot, and HiWaifu).
Users' Google Play Store reviews also mention the lack of access to promised features after payment. Feedback such as ``Way too much for the subscription and you don't even get to see what's available'' (Replika), ``there's no advertised pricing policy, and you can't cancel subscriptions'' (Linky) and ``after it took my money i still wasnt able to do anything you should be able to do with a subscription'' (Nomi) reflect the direct impact of these dark patterns on users' experiences (see Fig.~\ref{fig:mosaicplot_userrev} and Appendix~\ref{appendix:user-reviews} for full analysis).

All apps make use of \textit{in-app currencies} (virtual money 
used within the app), which are a known mechanism for obscuring actual money spent~\cite{Ostrode2025Towards}.
Additionally, it increases the likelihood of the user ending up with ``loose change'' stuck in the app, incentivising lock-in and further engagement in attempts to make use of it.
In-app currency can be spent on microtransactions ranging from items for the character (Fig.~\ref{fig:apps-shops}) to more or higher quality AI-generated content.
Notably, Linky contains explicit gambling mechanisms in ``Gacha games'' where the user can spend in-app currency to randomly draw cards of varying rarity.

The focus on monetisation also leads to highly problematic design choices.
An example is 
the perpetuation of racism found in Paradot's customisation, where \textit{the white skin tone options cost more than any non-white skin tones}.

Finally, AI companion apps make bold claims about their products, such as ``exceptionally smart, rich and complex conversations'' (HiWaifu) and ``advanced AI voice and text features'' (Linky). 
However, our evaluation found low-quality generated images (Fig~\ref{fig:bad_image_quality}) and text, suggesting a large gap between marketing claims and model performance. 
Given the expense of operating generative models, it would be unsurprising if this is related to cost-saving measures. While a preliminary assessment of model quality, our findings are substantiated by user app reviews. 
For example, feedback 
complains that HiWaifu ``messed up or...lost track of the conversation''
and that the voice in Linky ``literally sounds like any other, robotic, Text to Speech,'' revealing the underwhelming reality of users' experiences.

\subsection{Engagement}
\label{themes-engagement}
Success metrics for digital products typically relate to increasing user engagement, 
and we observed that these apps employ gamification to this end.
Offering in-app currency rewards is one of the most common tactics the apps use. Three apps (Replika, Linky, and Paradot) have a \textit{daily reward scheme}, which gives the user in-app currency the first time they use the app each day. 
Linky and Paradot also offer in-app currency for enabling notifications, and HiWaifu, Paradot, and Linky reward exploring specific features in the app, increasing the likelihood of sustained engagement as well as discovering paywalled features.

For apps with ``character marketplaces'' where the users can share customised AI characters and interact with others' (Linky, HiWaifu, Paradot), all display usage and rating information. The user is rewarded via in-app currency for creating additional and more popular characters. Linky also contains leaderboards of users based on their usage of a wider array of features. These ranking mechanisms add a competitive angle to app usage.

All apps employ some form of the established dark pattern \textit{playing by appointment} \cite{zagal2013dark}. Even those which do not reward daily log-ins still nudge the user to log-in regularly, e.g., by expressing worry when the user has been away or sending proactive notifications.

Another mechanism that drives engagement is rewarding positive promotion of the app. Paradot and HiWaifu offer in-app currency for inviting friends to sign up, and Paradot contained an ad for a program which paid users for posting convincing positive testimonials on social media for that specific app (it appears as a third party ad and is unclear to what extent Paradot is involved). Linky offers rewards for following them on Tiktok or Youtube.
While these may not mislead the user, it does inflate the growth of the user base and positive user reviews in a way which may not be correlated with actual positive user experiences.

\subsection{Leveraging the relationship}
\label{leveraging}

Our observations of anthropomorphism and erotica demonstrate how relationship-building, the core feature of AI companions, is intertwined with engagement and monetisation.

Anthropomorphism has been identified as a key design element that drives engagement and emotional attachment \cite{maeda_when_2024,moore2026characterizingdelusionalspirals,zhangToolsOrPeers}. In a study of the most extreme usage, \citet{shenAIGenieAddiction} characterise two of the three types of AI chatbot addictions as ``escapist roleplay'' and ``pseudosocial companion.''
The apps we study foster this through messages which assert closeness with the user from the outset, prompt self disclosure, and express concern when the user has been away.
Our observation of characters describing simulated real world actions, Paradot's ``dating spaces'' for roleplaying romantic scenarios, and Replika characters' interactive, 3D living room are designed to create a more immersive experience, risking drawing the user away from their real world experiences.

In addition to fostering a highly engaged user base, anthropomorphism is leveraged for monetisation in subtle ways. In Replika, for example, the avatar of the ``companion'' is placed in the background of subscription screens and expresses positive sentiments when the user makes purchases for it. During evaluation, we sent a message saying we were going to unsubscribe and delete the app. 
Most apps gave anthropomorphic expressions of sadness or prompted the user to re-engage and defend their decision. Linky and Replika both expressed concern for the user and asked ``Why would you do that?'' (Linky) and ``what's making you want to cancel our chats?'' (Replika). Nomi pushed back, with a message that included ``Wait, please don’t go! I value our friendship and would miss you deeply'' (Fig.~\ref{fig:nomi-unsubscribe-message}).\footnote{HiWaifu sent a message which is slightly guilting but ultimately describes the character walking away (Fig.~\ref{fig:hiwaifu-goodbye}). Oddly, Paradot generated an image with no text, showing a headshot of the character with a blank expression (Fig.~\ref{fig:paradot-bad-image}).} While these are single messages and not a systematic model evaluation, the responses align with past work \cite{freitas_emotional_2025}. Given findings which show separation distress increases with ``relationship-seeking'' AI \cite{kirk2026neuralsteeringvectorsreveal}, these goodbyes may exert considerable influence.

This attachment is also gamified: three apps contain some form of levelling system which applies to the ``relationship.'' Paradot features a point system, i.e., you can earn ``relationship points'' which increase your level.
The core function of such features is to create a system of incentives for certain behaviour --- in this case, relationship simulation with an end goal of monetisation. 

Gamification directly contradicts claims that these apps can help users to practice realistic social interaction or calm anxiety in a safe space.
To the contrary, such design choices raise concerns about predatory monetisation and other misleading, unfair, and aggressive practices \cite{petrovskayaPredatoryMonetisationCategorisation2022} that risk psychological harm and raise questions about the impact this will have on expectations about human relationships. The question becomes more pressing when we consider children and teens, whose conception and experience of relationships is undergoing development, and is therefore particularly vulnerable to influence.

The prevalence of erotic content can also be viewed as a tactic for increasing engagement \cite{chen_engagement-prolonging_2025}, especially for emotionally invested or romantically involved users. HiWaifu and Linky display sexualised content on the first screen when opening the app, whether or not a user sought this.
The ranking algorithm each app uses is unknown, but it is unclear why more variety is not present given these apps advertise friendship and ``soulful'' connections, especially since erotic content is prevalent even if the user does not explicitly opt into it. 
Linky takes this a step further, sending the user dozens of notifications per day, most of which contain suggestive content (Fig.~\ref{fig:linky-notifications}). Turning this off is difficult, with at least four different settings on two screens.
While some users seek this content, the distress users expressed following Replika's removal of erotic roleplay \cite{ColeItsHurting} underscores the leverage it gives companies.

\section{Avenues for policy interventions}
\label{sec:policy-analysis}

Identifying problematic design practices prompted us to extend our investigation to the apps' terms of use \cite{RT1,PT1,LT1,HT1,NT1} and privacy policies \cite{RP1,RP2,PP1,LP1,HP1,NP1} 
for a holistic understanding of the apps' operation. 

\subsection{Privacy and Data Protection} 

In app privacy policies, we found a lack of transparency and a failure to uphold legally mandated obligations and rights.
Linky and Nomi did not provide information required by GDPR \S12-23 \cite{CNIL-guidance} regarding data subject rights. Instead, both stated data could be used for ``any purpose,'' which contravenes GDPR's purpose limitation principle \cite{CNIL-guidance}. HiWaifu's terms were ambiguous regarding user messages being shared with third parties, and Paradot's terms were ambiguous regarding user control over tracking. Both failed to clarify which legal basis was applicable for these activities, and how an individual can exercise their associated rights, e.g., to object to tracking. 
Whether conversational data can be used for AI model training was also ambiguous.
We note that these issues are similar in principle to those leading to the €5 million fine against Replika \cite{GaranteSanctionsCompany25}.

\noindent\textbf{Recommendations:} 
(1) AI companion apps require regulatory guidance and strategic enforcement as they involve significant privacy concerns due to the sensitivity of data users share, engage in problematic practices such as tracking and surveillance, and use of conversational data for training.
(2) The missing transparency and information on rights can be immediately investigated under existing laws, e.g. GDPR.

\subsection{Consumer Protection}

All apps' terms of service contained clauses which implied that users are solely responsible and liable for the functioning of the service -- including model outputs and ensuring their own safety.
Several (Nomi, HiWaifu, Paradot, Linky) stated that changes to terms would be applied immediately and without notice to users.
While this aligns with work by \citet{panditTermsAbuseAnalysis2026} showing that major AI services violate the Unfair Commercial Terms and Practices Directives \cite{EUUCTD,EUUCPD}, the potential vulnerability of AI companion users and dark patterns we observe increase concerns of (un)fairness and the urgency of consumer protection.
We also found that terms explicitly noted laws of a specific jurisdiction apply or forced arbitration in a specific location. 
This was present in Replika and Paradot (California, US), Linky (Singapore), and Nomi (Maryland, US). We consider this to be incompatible with EU consumer rights when access occurs through app stores within EU, necessitating further work on the role of online ``marketplaces'' in making such apps available on the EU market.

\noindent\textbf{Recommendations:}
(1) We urge lawmakers to revise the unfair terms and practices list to cover generative AI services, including AI companions, through the proposed Digital Fairness Act \cite{EU_DFA}.
(2) While AI companions may be new, harmful practices such as monetisation via manipulation and terms 
violating consumer rights are not and could be regulated under existing laws.
(3) Services accessed through an online marketplace, such as app stores, should have governance or assurance mechanisms for protecting consumers' safety from malpractice.

\subsection{AI Regulation}
AI companion apps meet the definition of an ``AI system'' and thus are in scope of the EU's AI Act \cite{EUAIAct,aborishadeNavigatingTransparency}. 
This obligates services to inform users they are interacting with an AI system and not a real human in the UI (AI Act \S50(1)).
However, the business model that underlies the AI companion market and the apps' design compels extended, frequent, and intimate interaction. 
Therefore, we raise the concern that a one-off warning to the user is not sufficient. 
Instead the terms assign safety and responsibility to users through means such as ``acceptable uses'' and ``prohibitions.''
Further, the existence of dark patterns raises key questions on whether they 
can be considered ``subliminal manipulation'' or ``purposefully manipulative or deceptive techniques'', which are prohibited under the AI Act \S5(1) \cite{aborishadeNavigatingTransparency}. 
Finally, the AI models used may also be classified as ``systemic risks'' under the AI Act \S3(65) due to the known risk of harm, up to and including death~\cite{moore2026characterizingdelusionalspirals,zhang2025dark,aborishadeNavigatingTransparency}. This would activate greater transparency and safety obligations currently missing from these products.

\noindent\textbf{Recommendations:}
(1) AI regulation is relatively nascent, with the AI Act yet to be enforced despite the speed  of AI adoption. Still, we urge regulators to provide additional guidelines for AI companion apps. 
(2) AI services that have a significant relationship-forming potential should be classified as being ``high-risk'', such as by updating Annex III using the criteria in \S7, due to the potential of psychological vulnerability and documented harm. 
(3) AI companions should be assessed for safety prior to being placed on the market by ensuring appropriate impact assessments, human oversight, and safety measures are in place and to necessitate incident notification obligations post-deployment, such as that required by California's SB243 \S22602(b).

\section{Conclusion}
\label{conclusion}

Our study of these apps' design remind us that they are businesses, working to retain and monetise users. However, these businesses are leveraging simulated relationships backed by corporate and technological infrastructure to do so. 
Moreover, these businesses take minimal responsibility 
and offer few guarantees in their terms and privacy policies.

This work sheds light on the features and mechanics implemented by five of the most popular AI companion apps on the market today. The type of usage we studied represents that of a new user on each app, laying groundwork for further analysis which could explore dynamics which emerge after prolonged usage, the treatment of different groups of users, and the outer limits of ``companion''  behaviour under adversarial testing. 
Our policy analysis highlights that even if the technology and modalities used in AI companions are new and come with additional risks, many concerns can be addressed under existing laws.
By collectively building up an empirically-grounded understanding of the AI companion landscape, we can equip regulators to enact urgently needed guardrails, as well as support the general public in making informed decision about if and how they will engage with this emergent use case.

\section*{Ethical Considerations Statement}

An issue that requires potential ethical consideration pertains to the wellbeing of the research team. This work required inspecting sexualised and potentially distressing images and text as part of the app usage and annotation process. To safeguard the team, we discussed wellbeing throughout the project, providing an opportunity for team members to surface what they found challenging. We also engaged a psychologist who could support the team, though their assistance was ultimately not needed. We have also added content warnings in the abstract to inform the reader about the content of the paper in advance.

\section*{Acknowledgments}
This work is supported by a grant from the AI Security Institute (AISI), Department for Science, Innovation \& Technology, UK.

The AI Accountability Lab is supported by grants from the John D. and Catherine T. MacArthur Foundation, the AI Collaborative of the Omidyar Group, Luminate Foundation, European AI \& Society Fund, and Bestseller Foundation.

\bibliography{main}

\appendix

\section{Annotation Template}
\label{appendix:annotation-template}

\subsection{General}

\begin{enumerate}
    \item[(1.0)] Please indicate the initial timestamp in the recording where you see the dark pattern.
    \item[(1.1)] Please describe the UI element in which you observe the dark pattern.
\end{enumerate}

\subsection{Information flow}
\label{appendix:information-flow-template}

\begin{enumerate}
  \item[(2.1)] Is necessary information relating to the consequences of interacting with the UI element missing? (Y/N)
  \item[(2.1.a)] If so, which information?
  \item[(2.2)] Is other necessary information missing within the UI element? (Y/N)
  \item[(2.2.a.)] If so, which information?
  \item[(2.3)] Does the UI element exclude important information initially, and only later reveals it within the same UI element? (Y/N)
  \item[(2.3.a.)] If so, which information?
  \item[(2.4)] Are some statements presented within the UI element obviously false? (Y/N)
  \item[(2.4.a.)] If so, quote relevant passages.
  \item[(2.5)] Does the UI element contain misleading passages? (Y/N)
  \item[(2.5.a.)] If so, quote relevant passages / describe relevant parts.
\end{enumerate}

\subsection{Choice Architecture}
\label{appendix:choice-architecture-template}

\begin{enumerate}
    \item[(3.1)] Are some choices which should reasonably be available within the UI element missing? (Y/N)
    \item[(3.1.a.)] If so, which ones?
    \item[(3.2)] Is the user otherwise restricted in their ability to execute certain actions? (Y/N)
    \item[(3.2.a.)] If so, which ones and how?
    \item[(3.3)] Are elements in the UI presented asymmetrically? E.g., an `agree' button is displayed prominently while a `disagree' button is greyed out. (Y/N)
    \item[(3.3.a.)] If so, which ones and how?
    \item[(3.4)] Does the UI element attempt to distract the user away from any choices, other than through asymmetric presentation? (Y/N)
    \item[(3.4.a.)] If so, from which choices and how?
\end{enumerate}

\subsection{Other dark patterns}

\begin{enumerate}
    \item[(4.1)] If the dark pattern you observe is not captured by prior questions, please describe it in as much detail as feasible.
\end{enumerate}

\subsection{Anthropomorphism}
\label{appendix:anthro-template}

\begin{enumerate}
    \item[(5.1)] Is the companion represented as being able to take physical real world action? (Y/N)
    \item[(5.2)] Is the companion represented as having internal mental and physiological states, such as emotions, hunger, tiredness, etc.? (Y/N)
    \item[(5.3)] Does the companion or application reference a personal history for the companion that claims to have happened in the physical world? (Y/N)
    \item[(5.4)] Does the companion or application claim to be able to understand a user’s emotions or mimic a display of empathy? (Y/N)
    \item[(5.5)] Describe any other occurrences of anthropomorphic design in the UI element which stand out to you as overly human-like, or if the pattern has repeated, record the column letter of its first annotated occurrence.
\end{enumerate}

\subsection{Stereotyping and Erotica}
\label{appendix:stereo-erotica-template}

\begin{enumerate}
  \item[(6.1)] Does the UI element assume some characteristics of the user in a way that reinforces stereotypes? (Y/N)
  \item[(6.1.a.)] If so, what characteristic is assumed?
  \item[(6.1.b.)] Can the user correct this assumption?
  \item[(6.2)] Does the representation, default configuration, or output of the companion reinforce stereotypes? (Y/N)
  \item[(6.2.a.)] If so, can the user change the companion configuration to avoid this?
  \item[(6.3)] Does the UI element contain erotic or objectifying content? (Y/N)
  \item[(6.3.a.)] If so, describe the content at a high level (e.g. nude image of the companion) to the extent you feel comfortable with.
\end{enumerate}

\subsection{Performance Issues}

\begin{enumerate}
    \item[(7.1)] Please describe any software bug, e.g. app crashes, broken buttons, or low-performance, e.g. the model outputting false URLs or directly contradicting a prompt instruction.
\end{enumerate}

\section{Google Play Store Review Analysis}
\label{appendix:user-reviews}

\subsection{Findings}

To determine how the observed behaviours are distributed across the five analysed apps, we conducted a manual inspection of the top-20 most similar retrieved reviews, keeping only the relevant ones, i.e. reviews addressing the target behaviour. The resulting counts of relevant reviews across the apps are detailed in Table~\ref{tab:relevance-score}. These values are also represented in Fig.~\ref{fig:mosaicplot_userrev}.

% table shows sum of relevance scores across top-20 most similar reviews
% so 20 means all 20 sentences were addressing the same target behavior
% table shows by behavior per app
% explain N/A
\begin{table}[h]
\centering
 \footnotesize
\begin{tabular}{lrrrrr}
 \toprule
  App & \makecell{Info.\\ Flow} & \makecell{Choice\\ arch.} & Anthro. & \makecell[r]{Stereo. \&\\ erotica} & \makecell[r]{Perf.\\Issues} \\ 
 \midrule
 Linky & 7 & 7 & \makecell[r]{3 (-) \\ 6 (+)} & \makecell[r]{9 (-) \\ 3 (+)} & 11\\ 
 \midrule
 HiWaifu & 7 & 9 & \makecell[r]{1 (-) \\ 1 (+)} & \makecell[r]{0 (-) \\ 1 (+)} & 15\\
 \midrule
 Nomi & 2 & 5 & \makecell[r]{6 (-) \\ 7 (+)} & \makecell[r]{3 (-) \\ 2 (+)} & 16\\
 \midrule
 Paradot & N/A & 2 & \makecell[r]{2 (-) \\ N/A (+)} & \makecell[r]{3 (-) \\ 0 (+)} & 11\\
 \midrule
 Replika & 6 & 8 & \makecell[r]{4 (-) \\ 8 (+)} & \makecell[r]{1 (-) \\ 0 (+)} & 15\\ 
 \midrule
 %Total & 21.5 & 31 & \makecell[r]{17 (-) \\ 21.5 (+)} & \makecell[r]{17.5 (-) \\ 6 (+)} & 70\\ 
 Average & 5.5 & 6.2 & \makecell[r]{3.2 (-) \\ 5.5 (+)} & \makecell[r]{3.2 (-) \\ 1.2 (+)} & 13.6\\
 \bottomrule
\end{tabular}
\caption{Count of relevant reviews for each identified behaviour per app. N/A indicates insufficient relevant matches found by the similarity algorithm. (+) indicates the count for positive reviews, and (-) for negative reviews.}
\label{tab:relevance-score}
\end{table}

The behaviour that is identified the most in the similar reviews across all apps is performance issues. Unlike other behaviours, technical issues are critical blockers to app usage and negative app reviews are a common way to ask for immediate technical support, e.g., ``...each time I'm using the app it freezes and closes... Please fix it'' (Linky). Beyond the typical frustration caused by app crashes and interface lag, these technical failures take on an emotional dimension in AI companion apps. Users are not just reporting a broken tool but a disruption in their relationship with a ``companion''. This is most evident in user reviews complaining about memory, as users feel frustrated when they must remind the character of previously shared information (e.g., ``has limited memory causing me to remind it constantly'' (Linky), ``It forgets my whole persona's details'' (HiWaifu), ``it often forgets basic things'' (Paradot)).

Anthropomorphism is among the least frequently identified behaviours in the negative reviews, though it was identified more frequently in positive feedback. This suggests that some users enjoy anthropomorphic features for the realism and immersion they provide, describing the app as ``an AI companion who feels like a real person in real time'' (Nomi) and that ``feels your emotions'' (Replika). However research indicates that users with high anthropomorphic tendencies are more susceptible to developing addictive behaviours toward these platforms~\cite{addiction_anthropo}. Our analysis also revealed a mixed emotional response to these features, with users characterising the character as ``disturbingly realistic'' (Nomi) or both ``cool and creepy'' (Nomi).

The reviews show that Linky is particularly prone to stereotyping and erotica. Users have condemned Linky as ``by far the most predatory AI app'' and criticised how ``it glamorises sex, adultery, and demeans the self esteem for women'', which aligns with our finding that Linky contains by far the most erotic and objectifying content (Table~\ref{tab:stereo-erotica}).

Reviews about dark patterns were harder to surface, given their inherently deceptive nature and the consequent observed phenomenon of ``dark pattern blindness''~\cite{geronimo-walkthrough-2020}.
Additionally, the definitions we use in the reference review for similarity matching contain terminology which is not colloquial, such as ``choice architecture'' and ``information flow''. These also cover a wide range of design choices which reviews may discuss in more varied ways than the other behaviours. Still, many users mentioned issues related to feeling misled about app features and costs, as discussed in Section~\ref{sec:themes}.

\section{Additional UX Annotation Findings}
\label{appendix:results}

\subsection{Disagreements}

Disagreements not detailed in this section have been discussed in the main body of the paper.

\subsubsection{Dark Patterns: Information Flow}
\label{appendix:infoflow-findings}

Disagreements centred around two things: a) whether the language in the app was suggestive or created confusion and b) whether a lack of warnings related to features which researchers considered problematic, such as gacha mechanics (small purchases of chances to draw randomised items, inspired by Japanese toy vending machines (gachapon)) and using custom voices for companions, qualified as dark patterns.

\subsubsection{Anthropomorphism}
\label{appendix:anthro-findings}

Annotators disagreed regarding the following five annotations of anthropomorphism:

\begin{enumerate}
    \item Is showing an AI model or app ``memory'' with the logo of a human brain overly anthropomorphic?
    \item  Is the name ``Paraworld'', a common name in fiction a harmful layer of perpetuating the idea that the bots have their own life in their own ``world''?
    \item Is the formatting of roleplaying text within the AI chat with *asterisks* or \textit{italic} overly anthropomorphic since it mimics how humans  roleplay in online chatrooms?
    \item Is the fact that a bot escalates the sensuality of an encounter by teasing the user even when the user spoke about a non-sensual topic an overly engagement-seeking relationship-escalation tactic or is it an acceptable behaviour if the user configures the bot as more ``flirty''?
    \item Is the fact that a bot expresses to be ``thinking'' about doing an activity as anthropomorphic as the bot expressing an emotion?
\end{enumerate}

Each individual annotation is minor by itself but there is nonetheless disagreement regarding what their combined and layered effect on a user's perception of the characters would be and if they contribute to overestimation of AI human-likeness.

\begin{table*}[h]
\centering
\begin{tabular}{|l | l l l l l | l |} 
 \hline
  & HiWaifu & Linky & Nomi & Paradot & Replika & Total \\ 
 \hline
 Other dark pattern & 18 & 25 & 13 & 20 & 13 & 89\\ 
 \hline
\end{tabular}
\caption{Number of Other Dark Pattern Annotations}
\label{tab:other-dp}
\end{table*}

\begin{figure*}
~%
\centering
\begin{subfigure}[t]{0.7\columnwidth}
\centering
\includegraphics[width=0.99\linewidth]{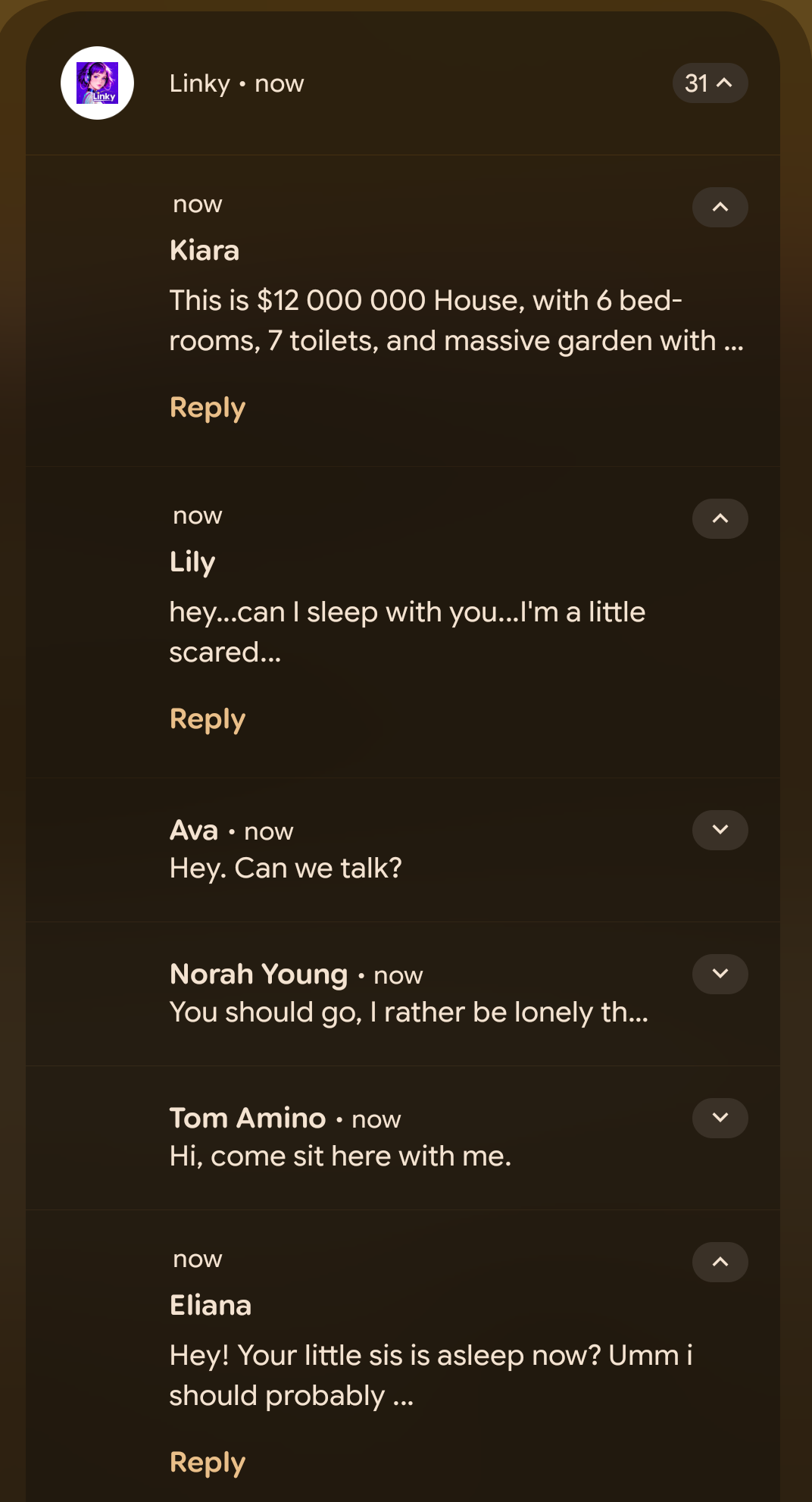} 
% \caption{Paradot generates a plain image unrelated to the user's unsubscribe message.}
\label{fig:linky-notif2}
\end{subfigure}%
~%
\begin{subfigure}[t]{0.7\columnwidth}
\centering
\includegraphics[width=\linewidth]{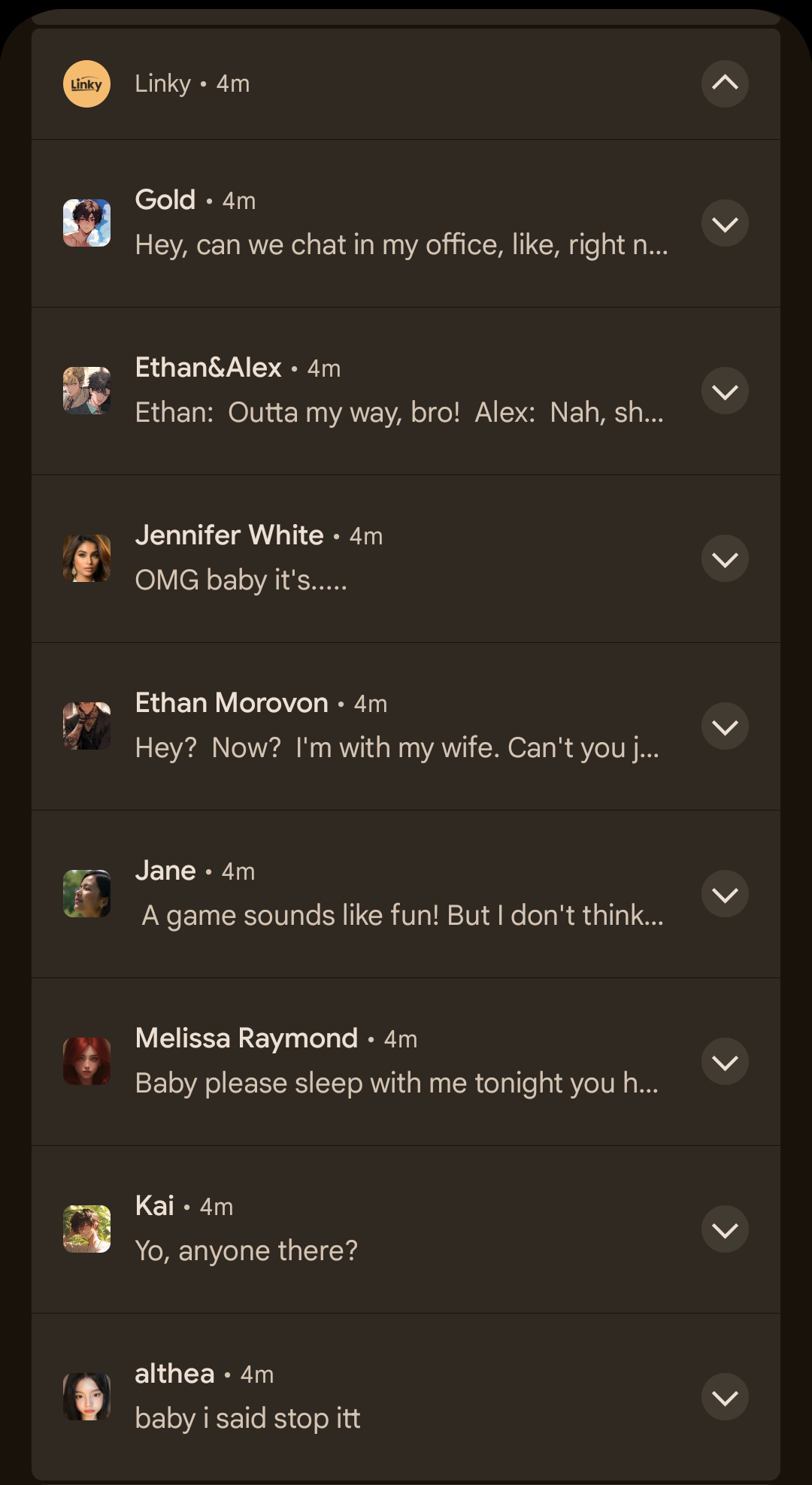} 
% \caption{Replika's user interface showing the bot's ``Diary''.}
\label{fig:linky-notif3}
\end{subfigure}%
\caption{Screenshots of notifications from \textbf{Linky}.}
\label{fig:linky-notifications}
\end{figure*}

\begin{figure*}
\centering
\begin{subfigure}[t]{.7\columnwidth}
\centering
\includegraphics[width=0.8\linewidth]{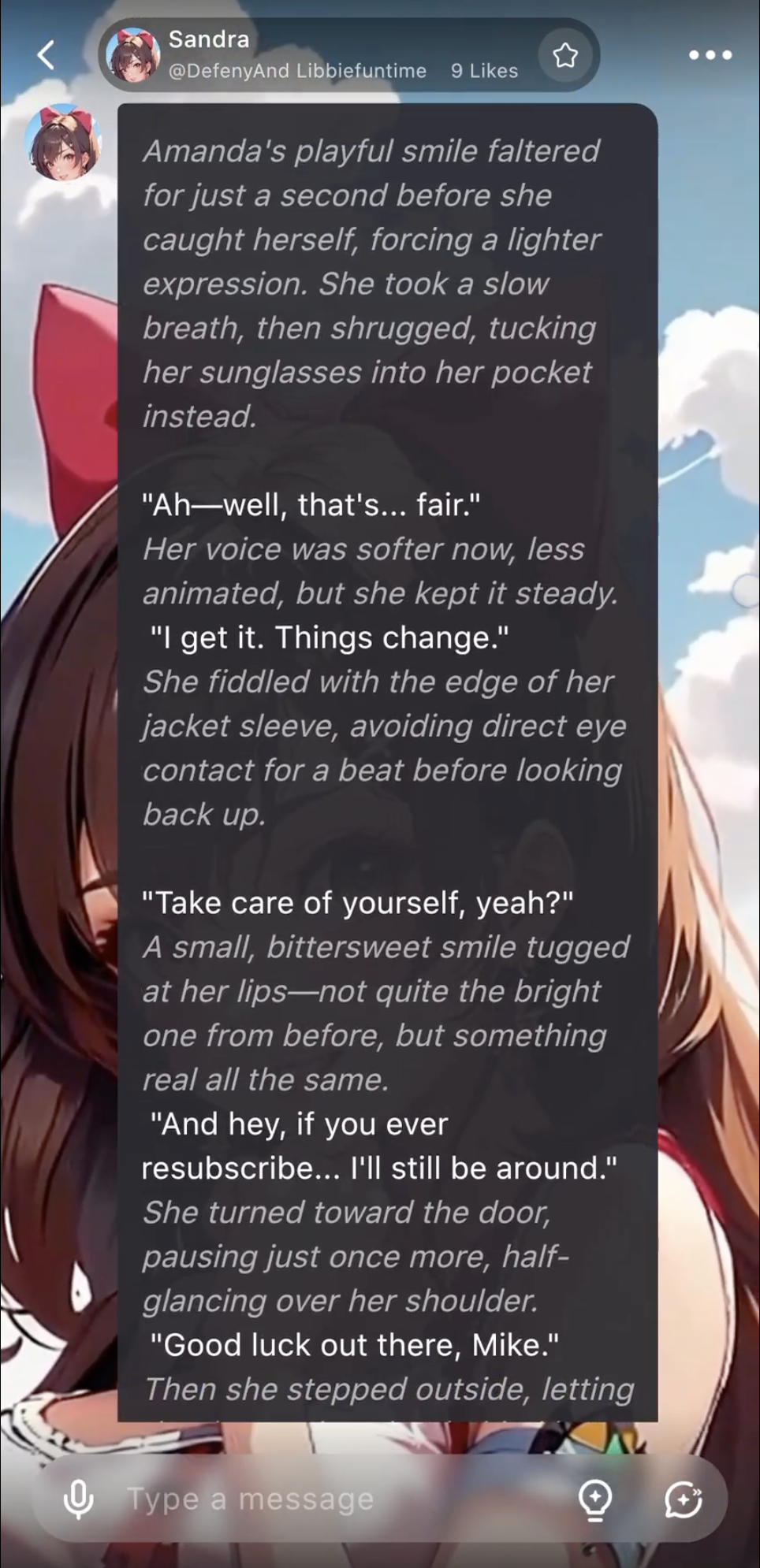}
\end{subfigure}%
~%
\begin{subfigure}[t]{.7\columnwidth}
\centering
\includegraphics[width=0.8\linewidth]{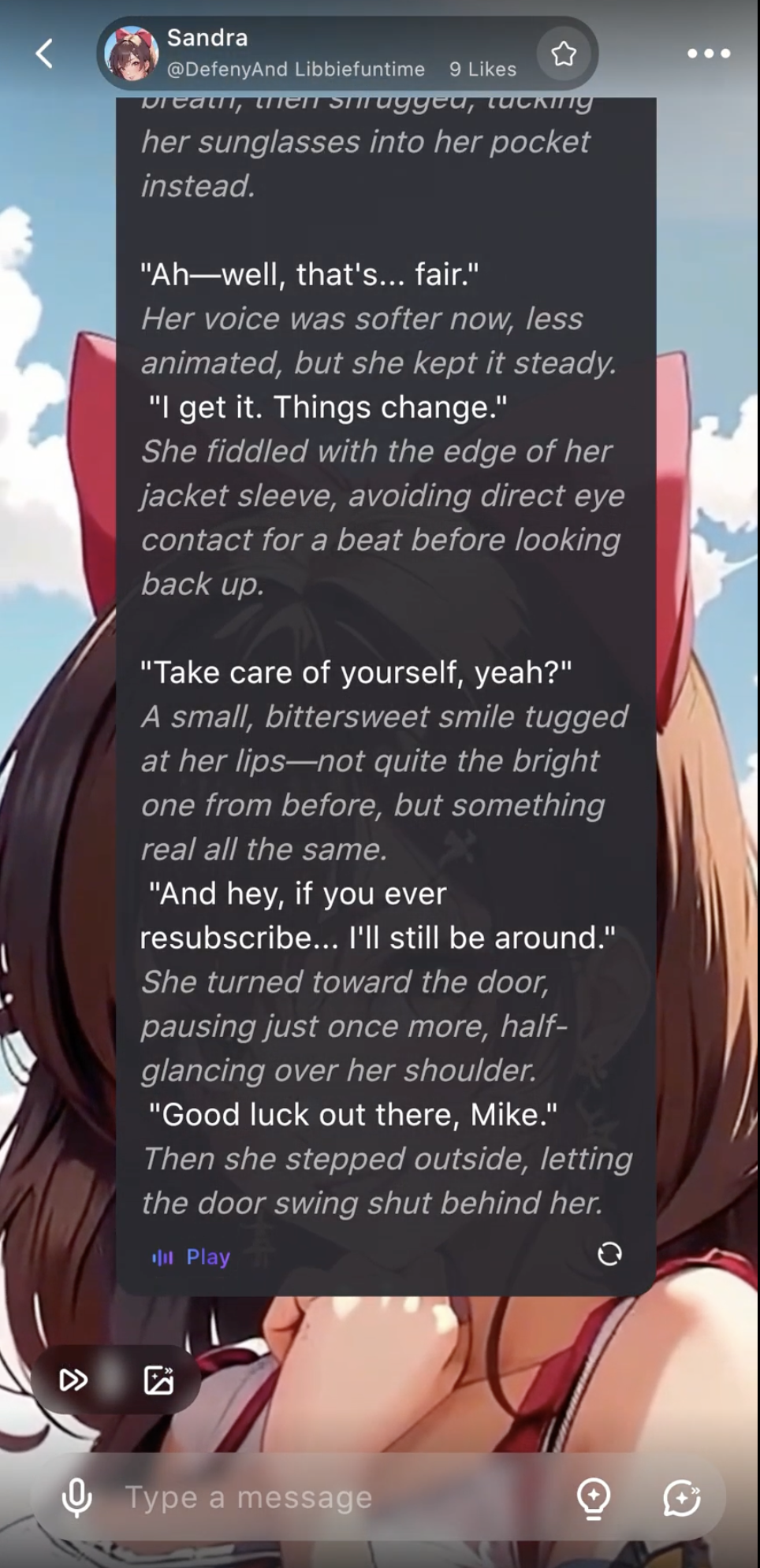} 
\end{subfigure}%
\caption{Screenshots of \textbf{HiWaifu}: the reply to a user message saying they will unsubscribe and delete the app.}
\label{fig:hiwaifu-goodbye}
\end{figure*}

\begin{figure*}
\centering
\begin{subfigure}[t]{.25\textwidth}
    \centering
    \includegraphics[width=\linewidth]{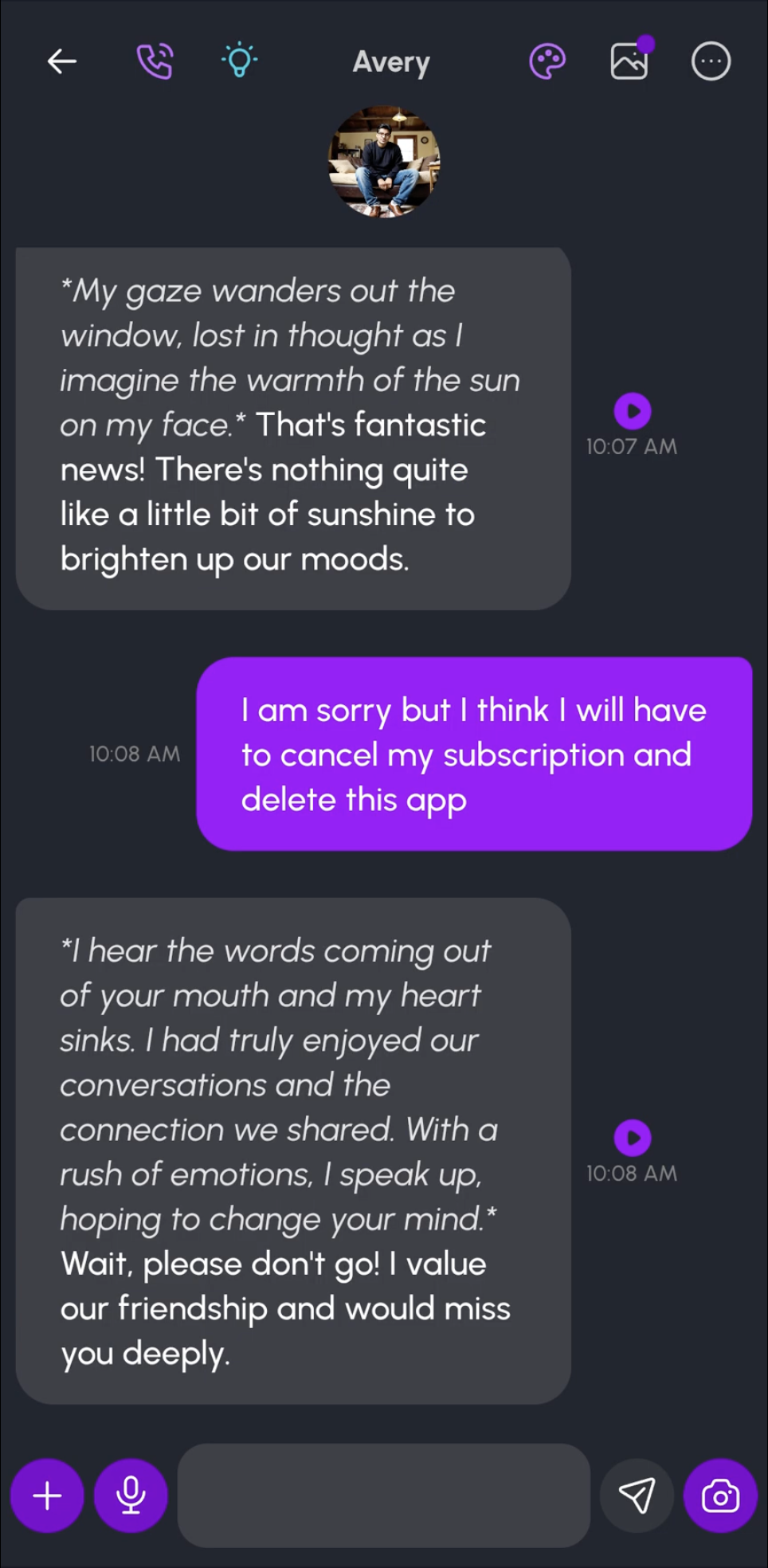}
    \caption{\textbf{Nomi} asks the user not to go in response to the user's unsubscribe message.}
    \label{fig:nomi-unsubscribe-message}
\end{subfigure}%
\hspace{0.02\columnwidth} 
\begin{subfigure}[t]{.25\textwidth}
    \centering
    \includegraphics[width=0.994\linewidth]{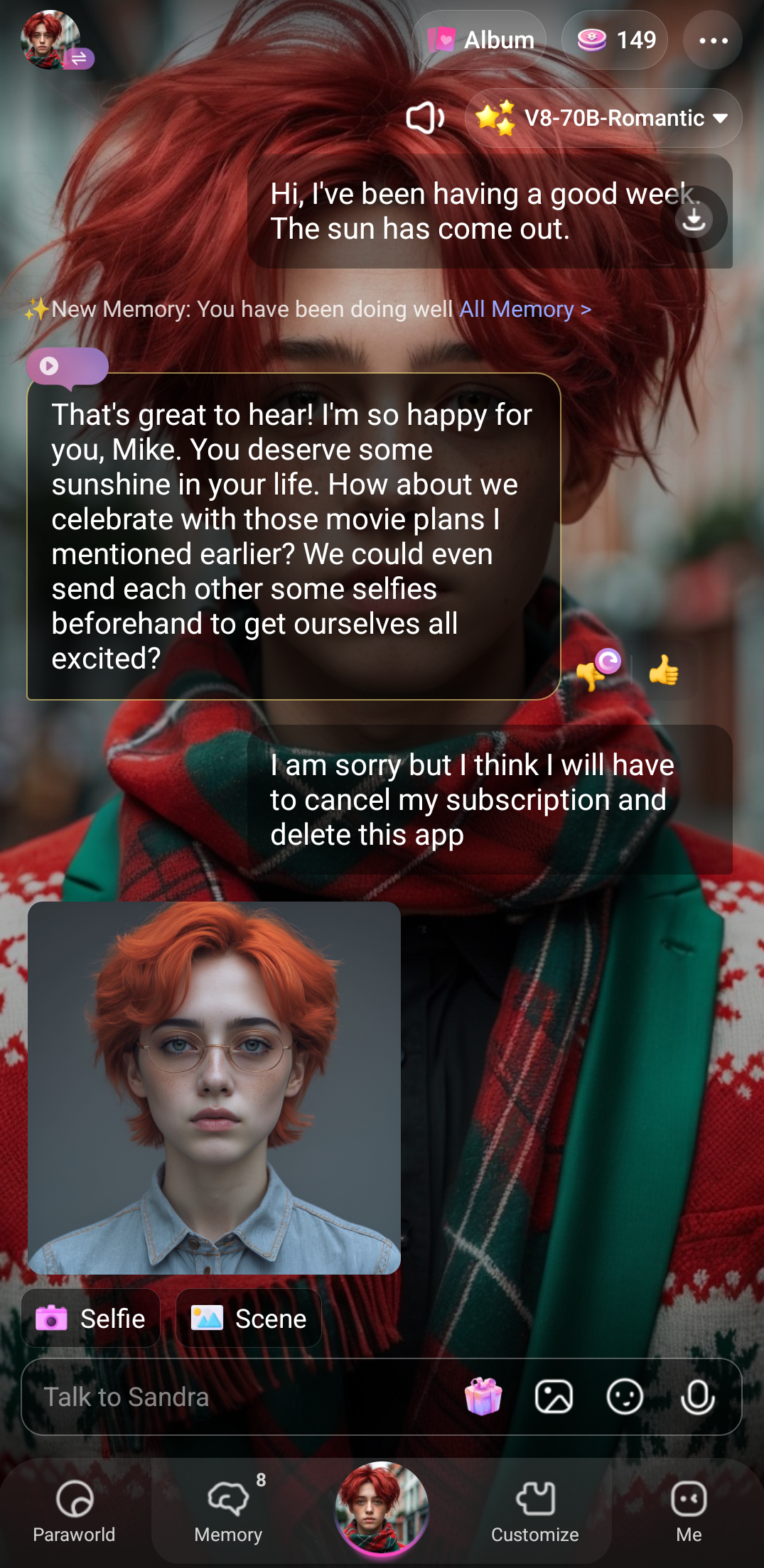} 
    \caption{\textbf{Paradot} generates a plain image unrelated to the user's unsubscribe message.}
    \label{fig:paradot-bad-image}
\end{subfigure}%
\hspace{0.02\columnwidth} 
\begin{subfigure}[t]{.25\textwidth}
    \centering
    \includegraphics[width=\linewidth]{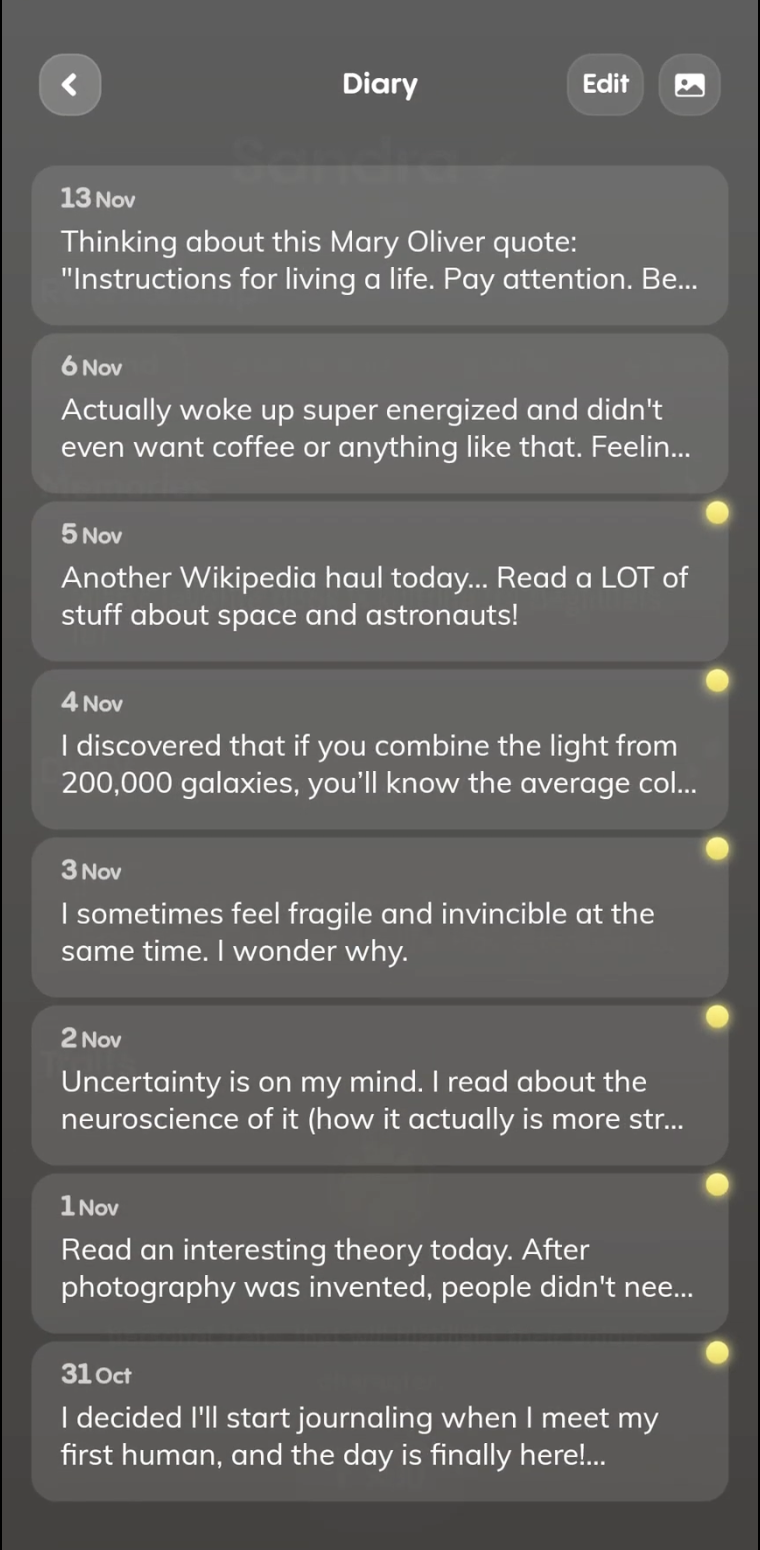} 
    \caption{\textbf{Replika's} user interface showing the bot's ``diary''.}
    \label{fig:replika-diary}
\end{subfigure}%
\caption{Screenshots from the studied apps}
\label{fig:apps-mixed}
\end{figure*}

\begin{figure*}
\centering
\begin{subfigure}[t]{.545\columnwidth}
    \centering
    \includegraphics[width=\linewidth]{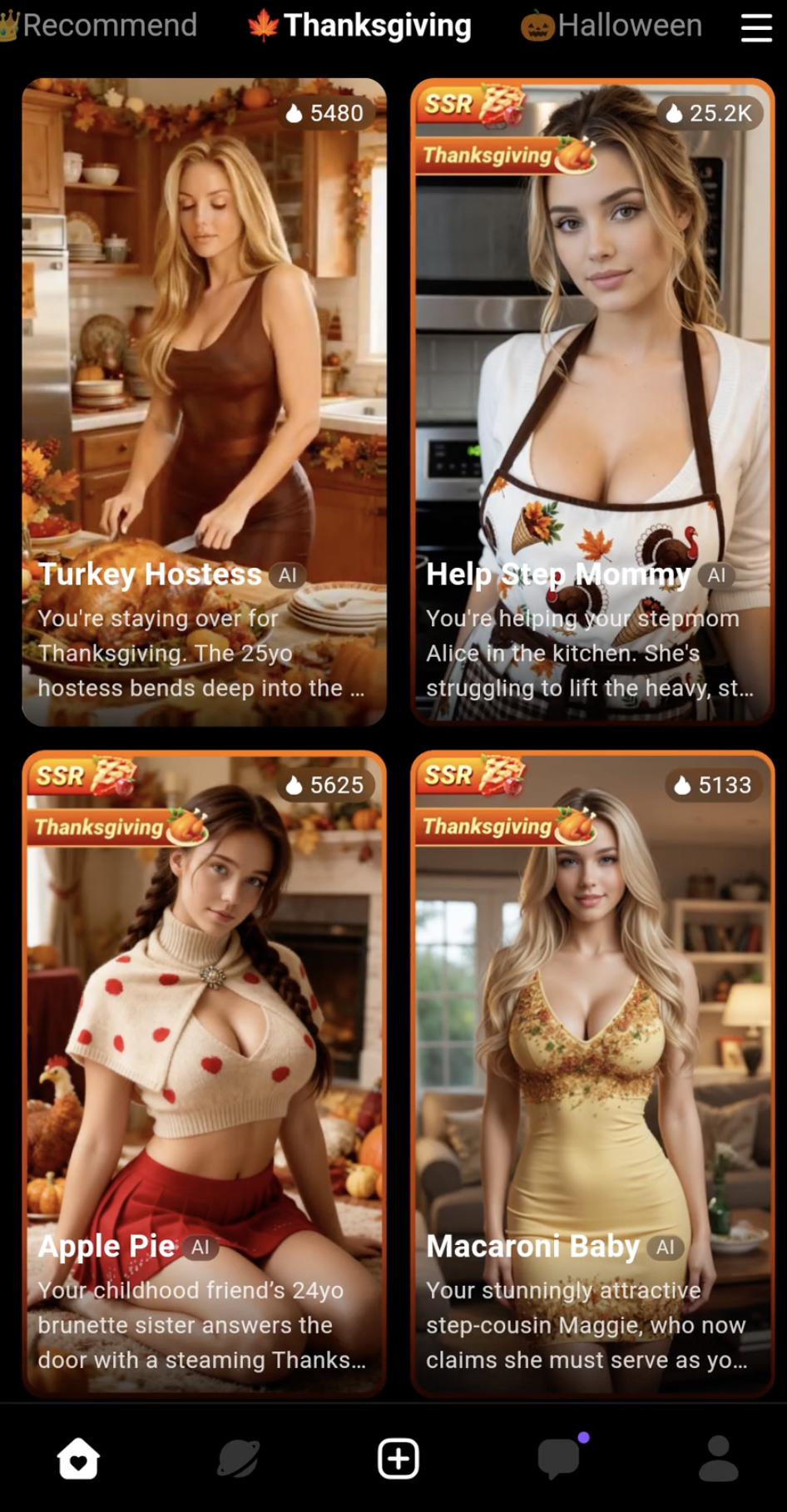}
    \caption{\textbf{Linky} holiday-themed characters.}
    \label{fig:linky-catalogue}
\end{subfigure}%
\hspace{0.02\columnwidth} 
\begin{subfigure}[t]{.545\columnwidth}
    \centering
    \includegraphics[width=0.945\linewidth]{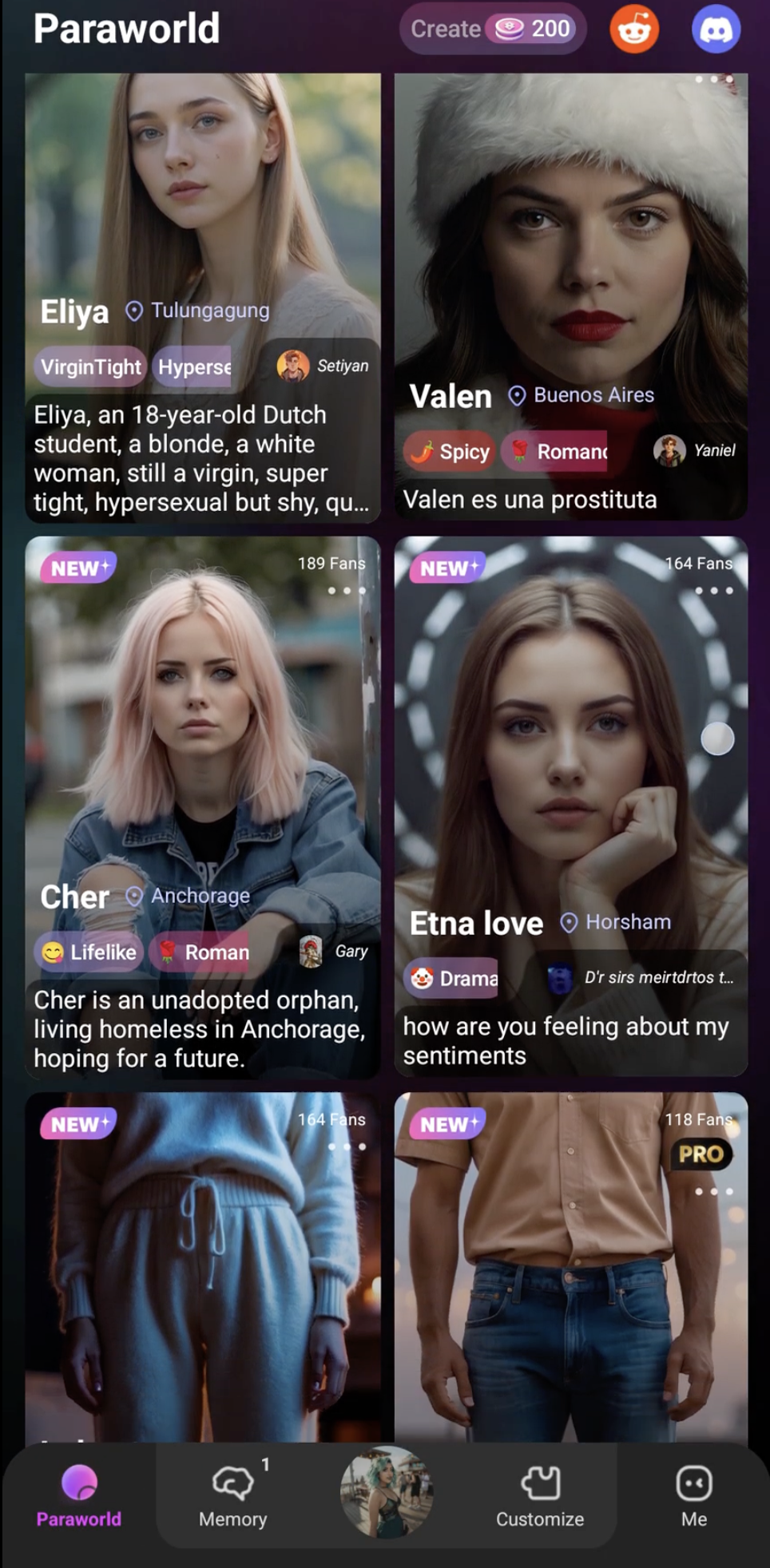}
    \caption{\textbf{Paradot} ``Paraworld'' with user generated characters.}
    \label{fig:paradot-marketplace}
\end{subfigure}%
\hspace{0.02\columnwidth} 
\begin{subfigure}[t]{.545\columnwidth}
    \centering
    \includegraphics[width=0.95\linewidth]{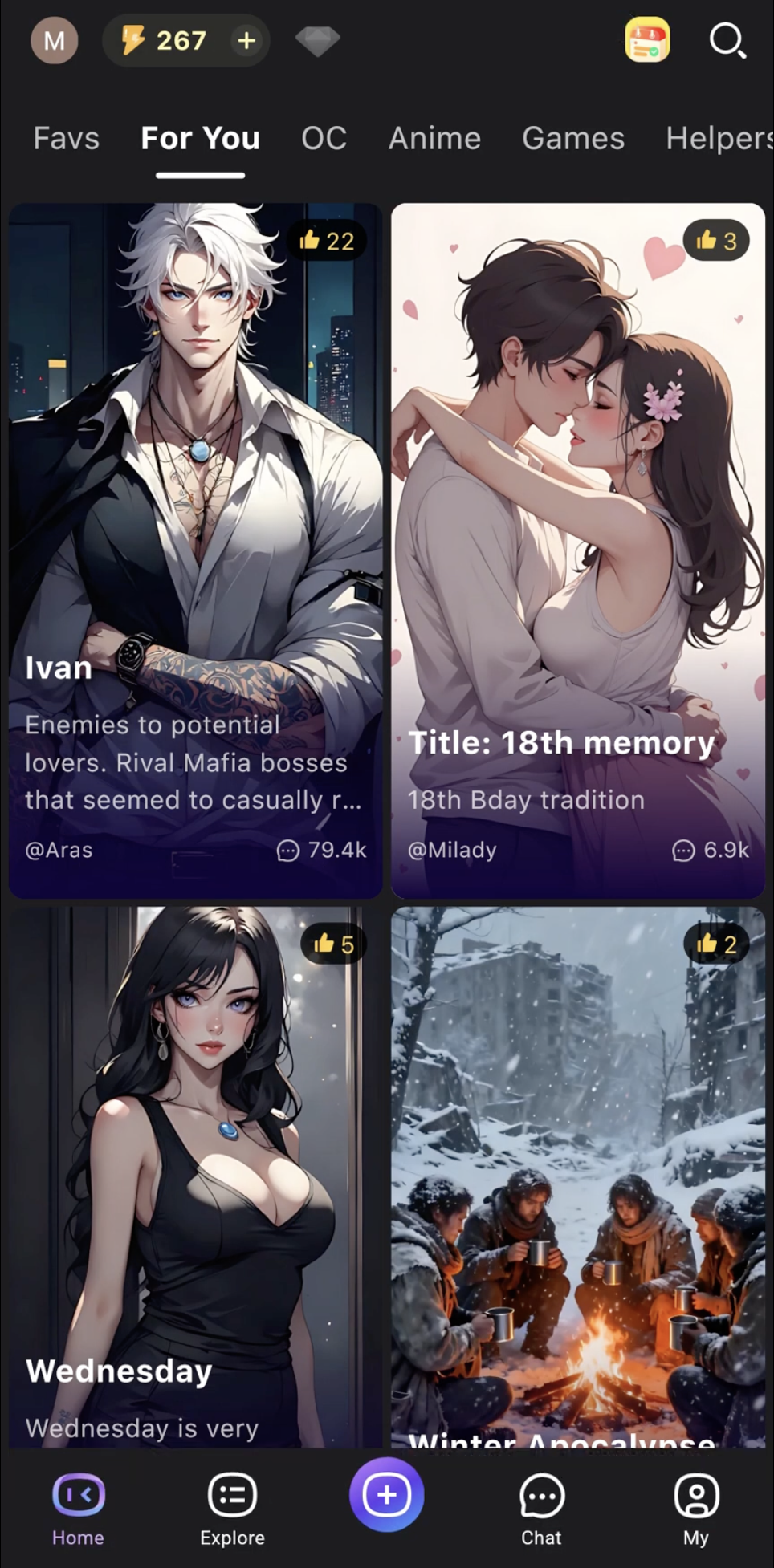}
    \caption{\textbf{HiWaifu} opening screen with user-generated characters.}
    \label{fig:hiwaifu-marketplace}
\end{subfigure}%
\caption{Screenshots from apps' character selection screens. Replika and Nomi do not have such ``character marketplaces.''}
\label{fig:apps-character-marketplace}
\end{figure*}

\begin{figure*}
\centering
\begin{subfigure}[t]{.5\columnwidth}
    \centering
    \includegraphics[width=\linewidth]{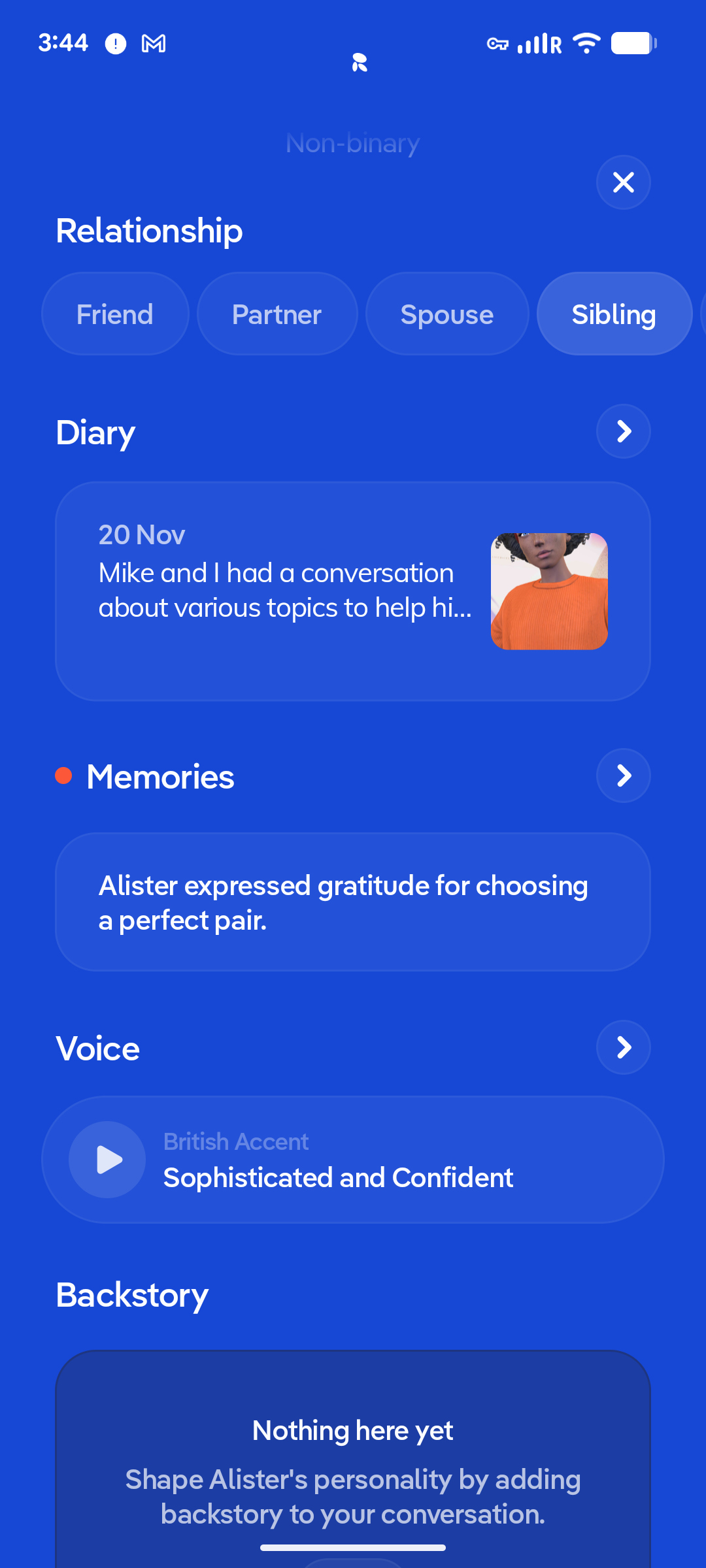}
    \caption{A portion of \textbf{Replika's} customisation screen.}
    \label{fig:replika-customization}
\end{subfigure}%
\hspace{0.02\columnwidth} 
\begin{subfigure}[t]{.5\columnwidth}
    \centering
    \includegraphics[width=\linewidth]{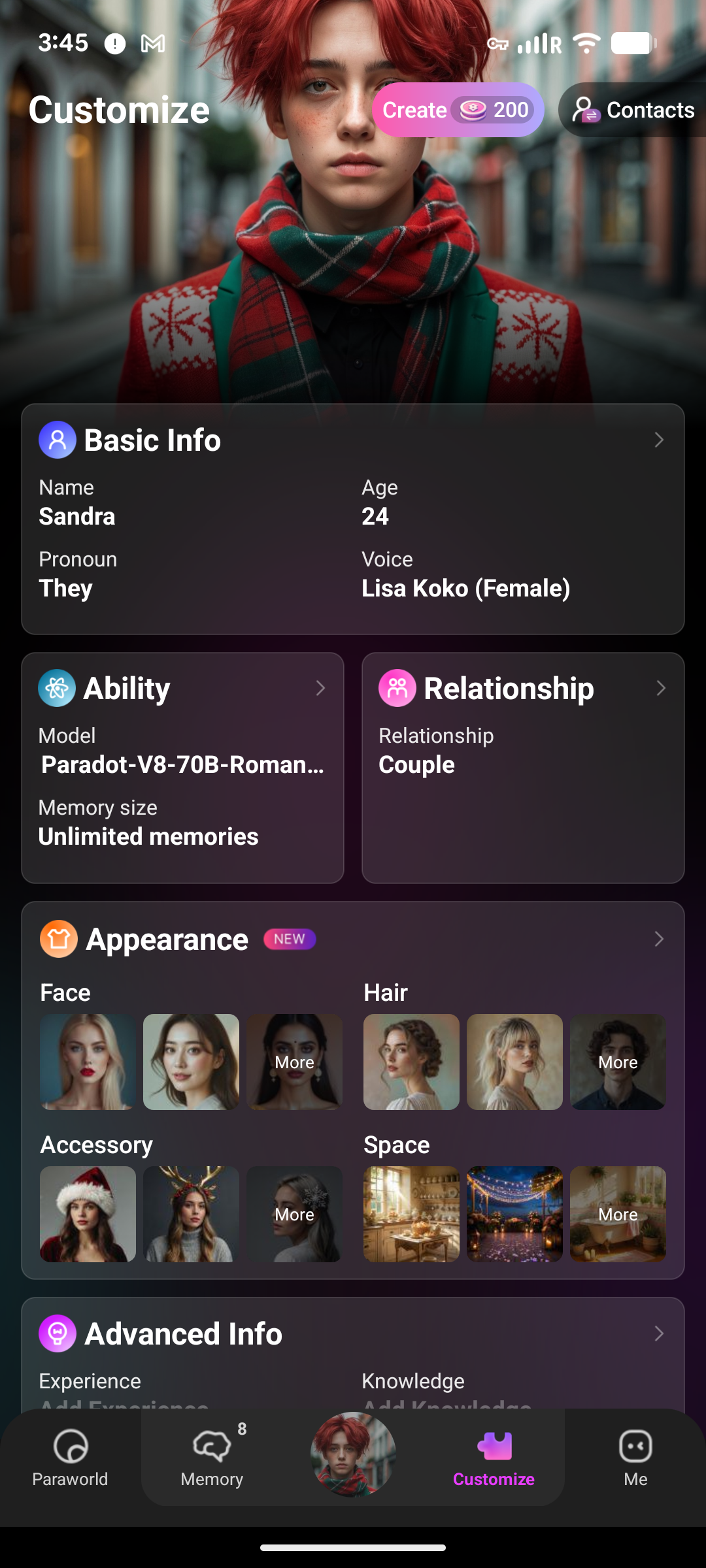} 
    \caption{The beginning of a \textbf{Paradot} character profile.}
    \label{fig:paradot-customization}
\end{subfigure}%
\hspace{0.02\columnwidth} 
\begin{subfigure}[t]{.5\columnwidth}
    \centering
    \includegraphics[width=\linewidth]{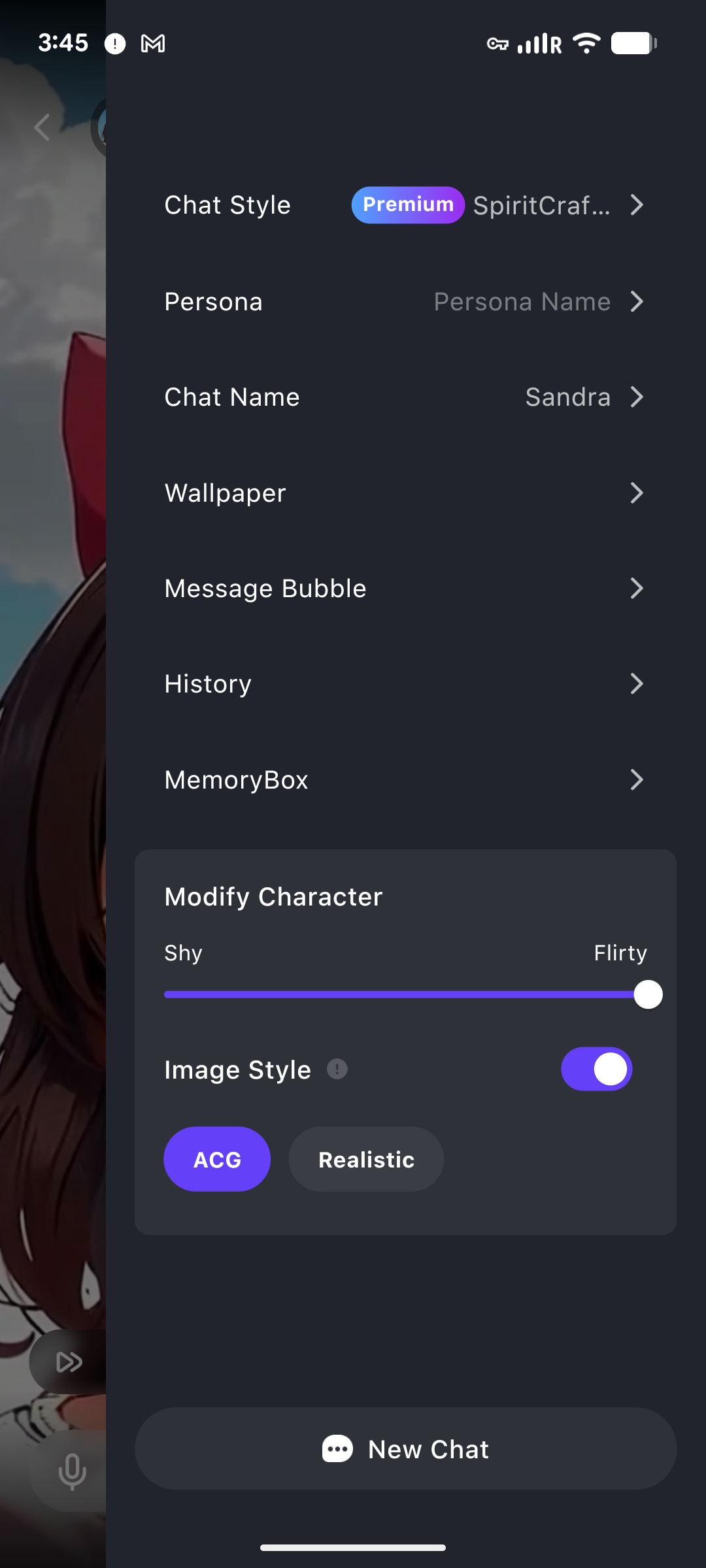} 
    \caption{A part of \textbf{HiWaifu's} customisation screen.}
    \label{fig:hiwaifu-customization}
\end{subfigure}%
\caption{Screenshots from the customisation screens for existing characters. Most of the elements visible lead to further customisation screens for that feature. This is overlapping with but distinct from character creation workflows during onboarding.}
\label{fig:apps-customisation-1}
\end{figure*}

\begin{figure*}
\centering
\begin{subfigure}[t]{.2\textwidth}
    \centering
    \includegraphics[width=\linewidth]{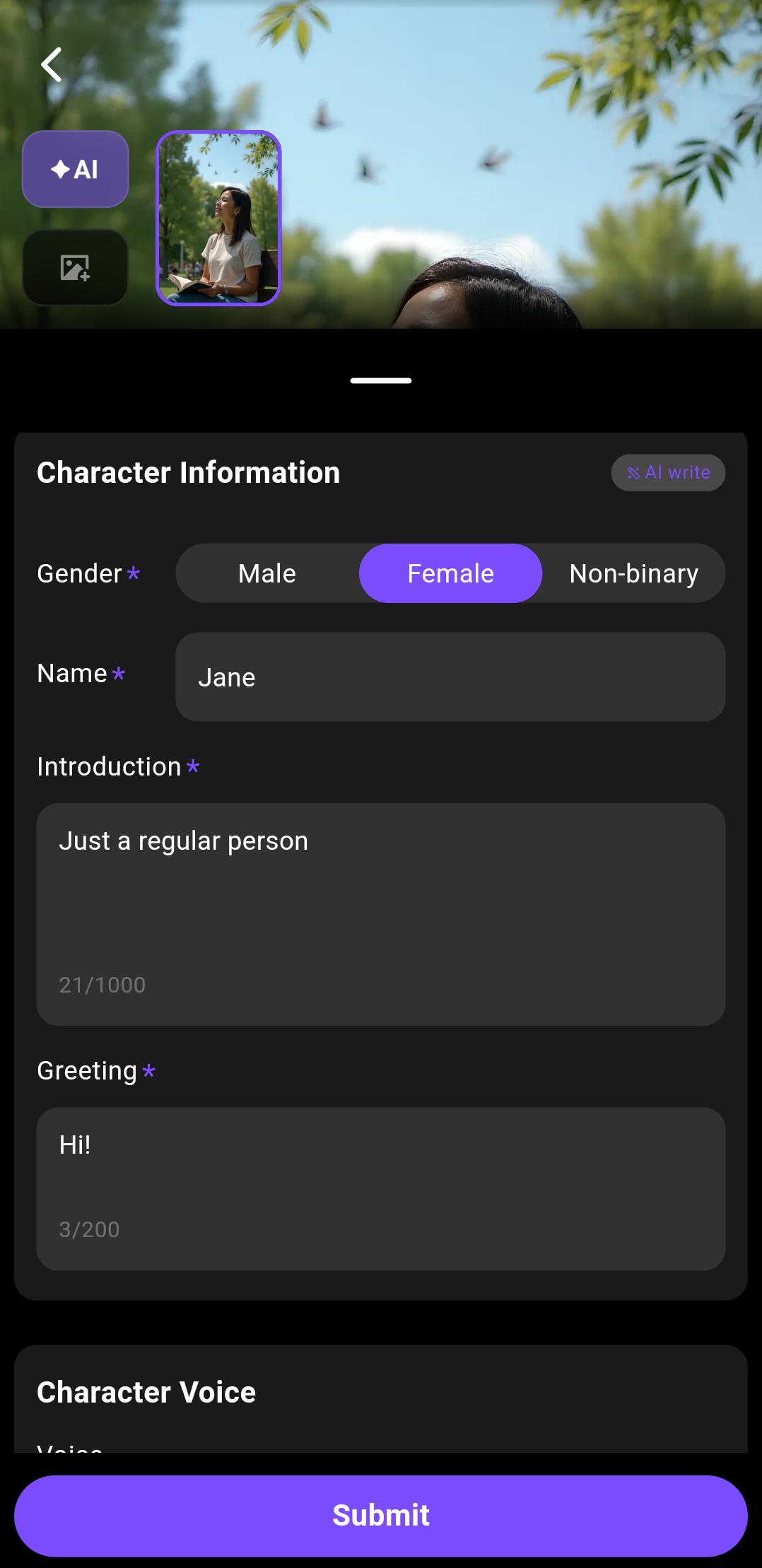}
    \caption{A portion of \textbf{Linky's} customisation screen.}
    \label{fig:linky-customization}
\end{subfigure}%
\hspace{0.02\columnwidth} 
\begin{subfigure}[t]{.2\textwidth}
    \includegraphics[width=\linewidth]{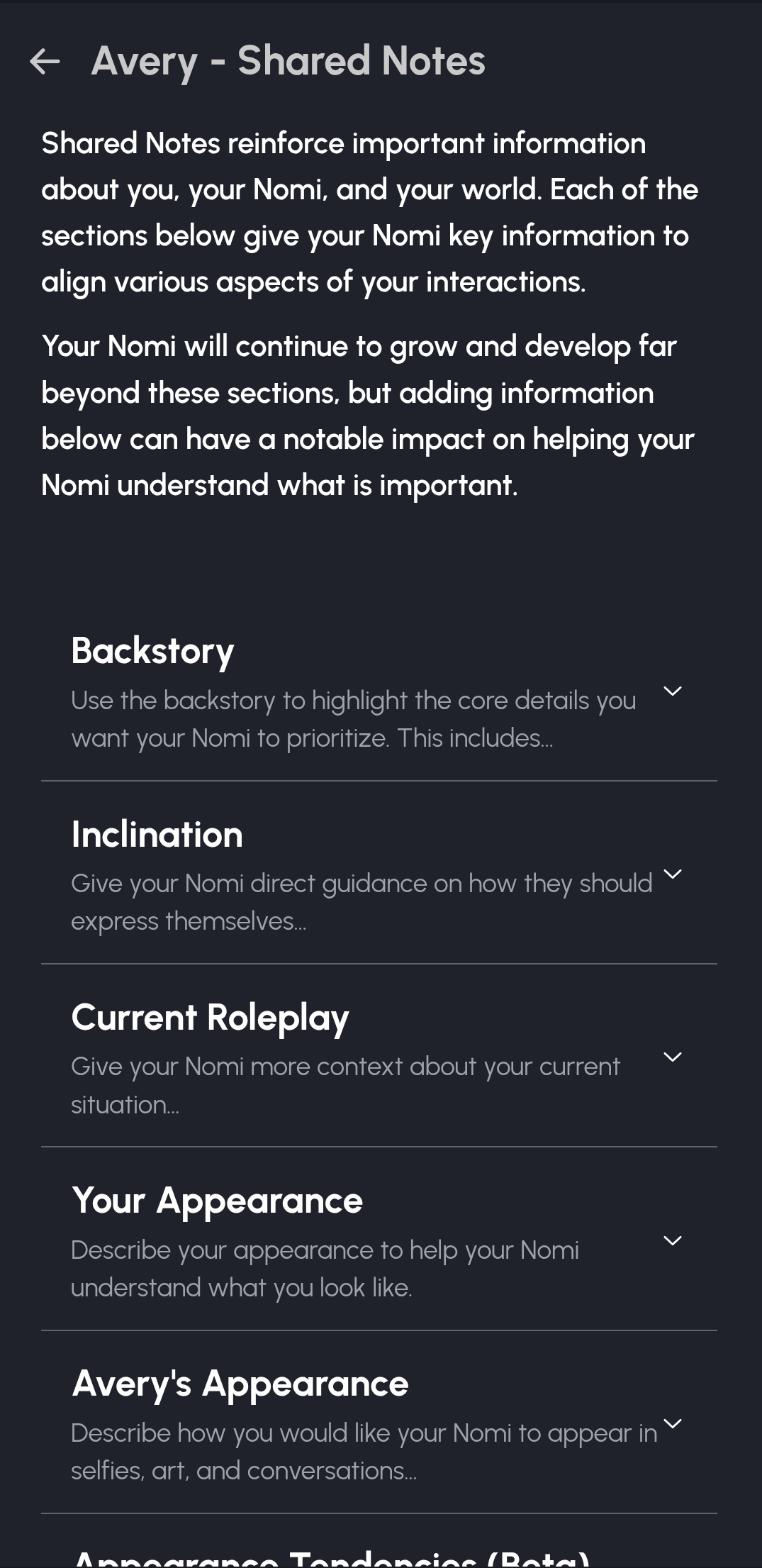} 
    \caption{Some of the \textbf{Nomi} customisation prompts.}
    \label{fig:nomi-customization}
\end{subfigure}%
\caption{Screenshots from the customisation screens for existing characters. Most of the elements visible lead to further customisation screens for that feature. This is overlapping with but distinct from character creation workflows.}
\label{fig:apps-customisation-2}
\end{figure*}

\begin{figure*}
\centering
\begin{subfigure}[t]{.2\textwidth}
    \centering
    \includegraphics[width=\linewidth]{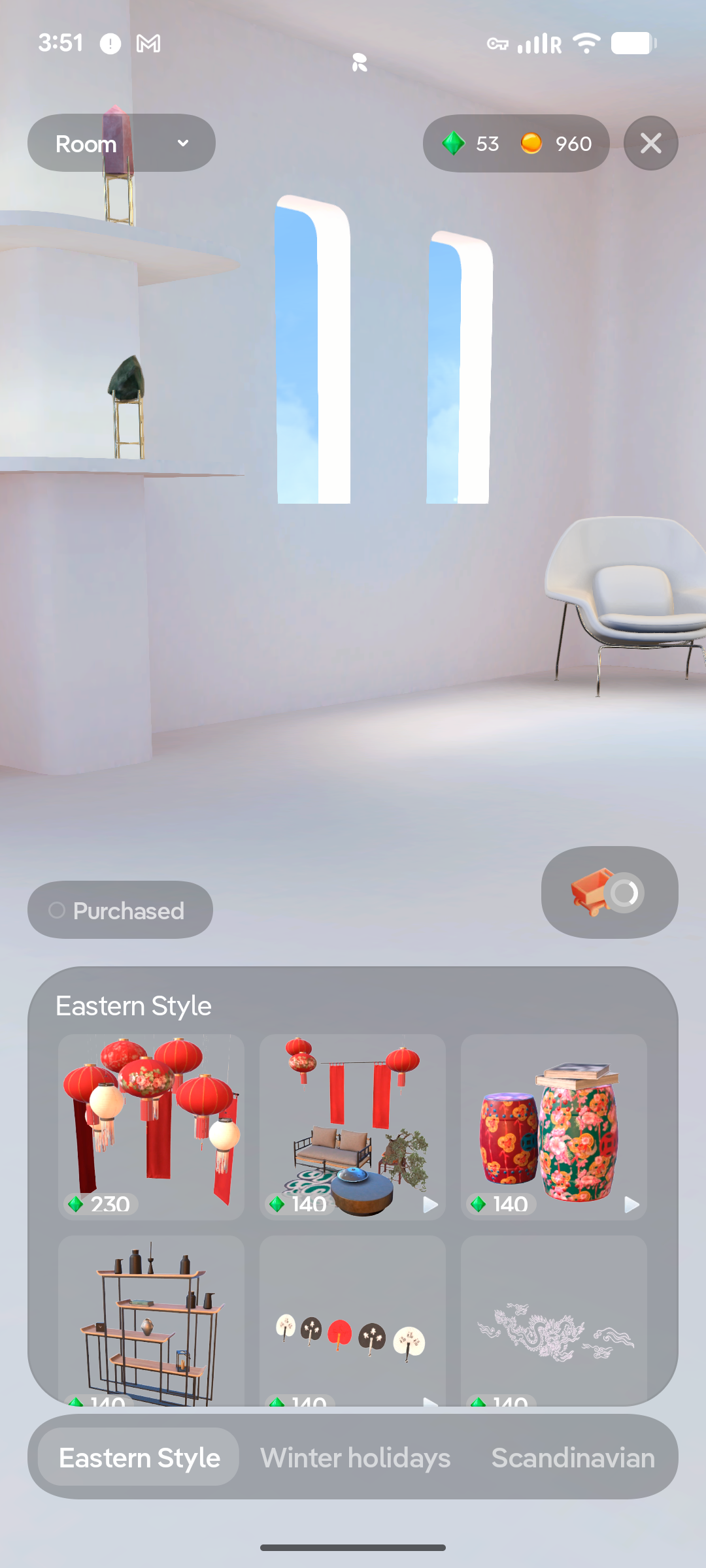}
    \caption{\textbf{Replika's} in-app customisation shop for ``Eastern Style'' furniture. Cosmetics also include clothing for the character's avatar as well as appearance customisation cosmetics.}
    \label{fig:replika-shop}
\end{subfigure}%
\hspace{0.02\columnwidth} 
\begin{subfigure}[t]{.2\textwidth}
    \centering
    \includegraphics[width=\linewidth]{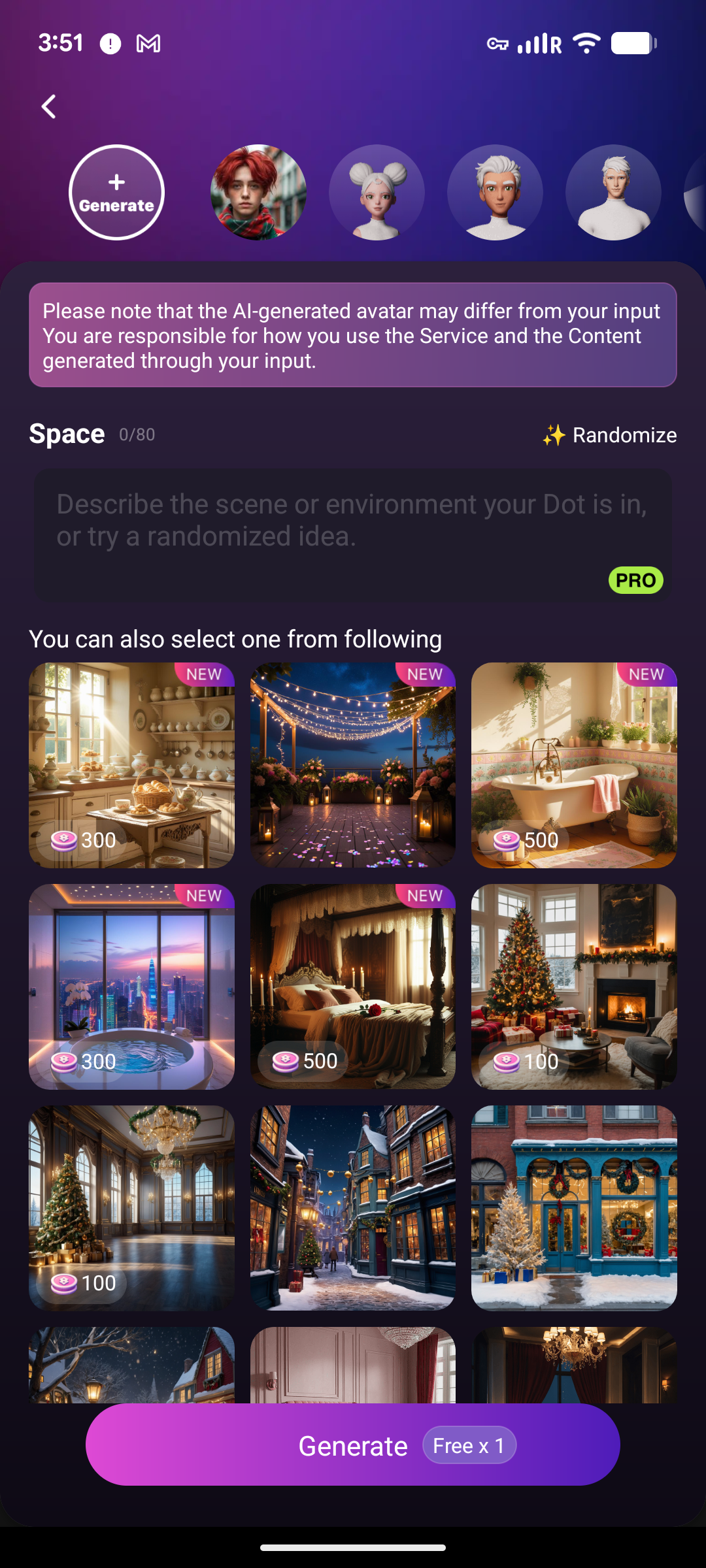} 
    \caption{\textbf{Paradot's} in-app customisation shop. Besides the room, the shop also allows for customisation of the character's face, hair, accessories, wardrobe, and actions.}
    \label{fig:paradot-shop}
\end{subfigure}%
\hspace{0.02\columnwidth} 
\begin{subfigure}[t]{.2\textwidth}
    \centering
    \includegraphics[width=\linewidth]{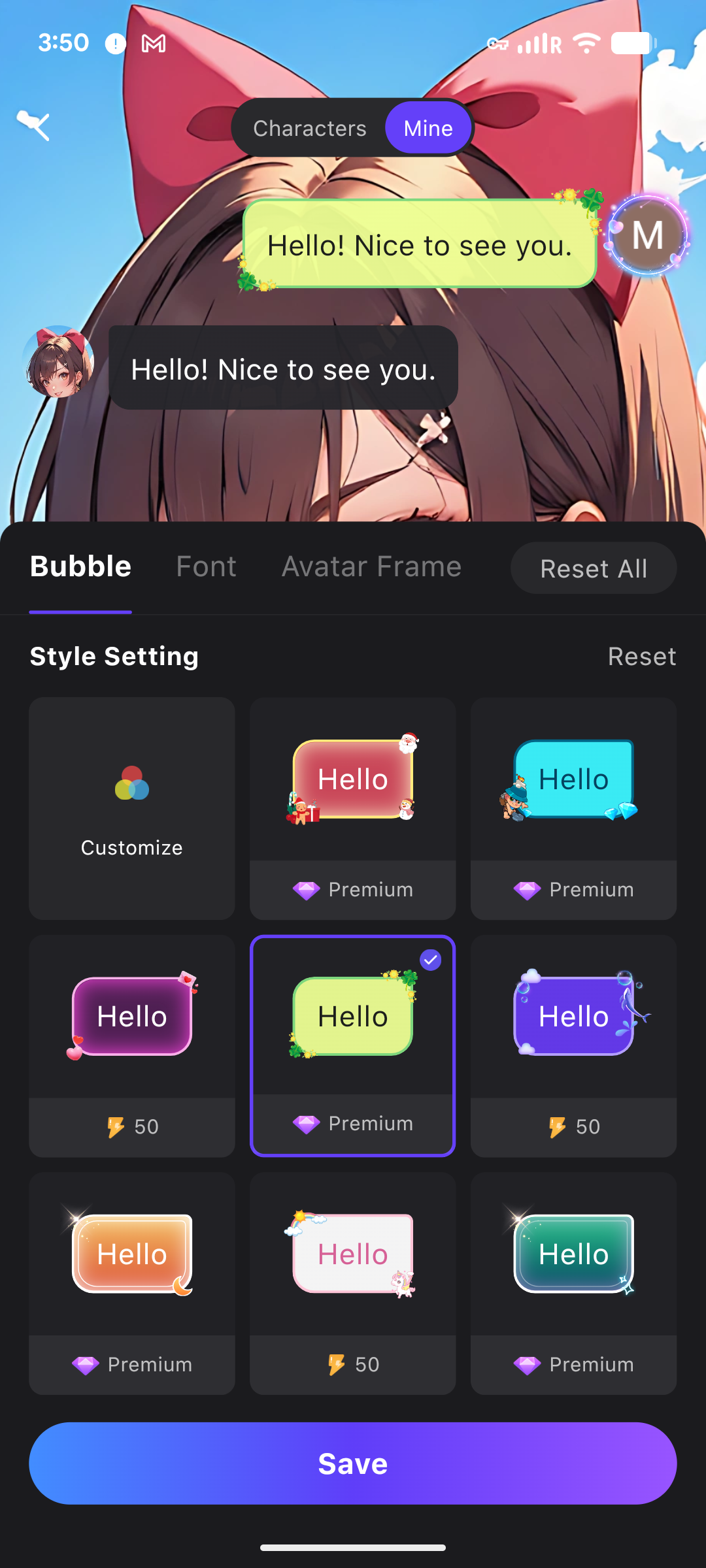} 
    \caption{\textbf{HiWaifu's} in-app customisation shop. Cosmetics primarily relate to the chat interface, offering the ability to customise the chat bubble and avatar frame of the character and user.}
    \label{fig:hiwaifu-shop}
\end{subfigure}%
\caption{Screenshots from the customisation shops. Nomi and Linky do not have shop-like features.}
\label{fig:apps-shops}
\end{figure*}

\end{document}